\documentclass[12pt]{cernalice}
\usepackage[english]{babel}
\usepackage[dvips]{graphicx}
\usepackage{verbatim}
\usepackage{makeidx}
\usepackage{color}
\newcommand{\av}[1]{\left\langle #1 \right\rangle}
\newcommand{\sqrtsNN}{\sqrt{s_{\scriptscriptstyle{{\rm NN}}}}}
\newcommand{\PbPb}{\mbox{Pb--Pb}}
\newcommand{\pPb}{\mbox{p--Pb}}

\newcommand{\NN}{\mbox{nucleon--nucleon}}
\newcommand{\pp}{\mbox{proton--proton}}
\newcommand{\pA}{\mbox{proton--nucleus}}
\renewcommand{\AA}{\mbox{nucleus--nucleus}}
\newcommand{\pt}{\ensuremath{p_{\mathrm{t}}}}

\newcommand{\fm}{\mbox{${\rm fm}$}}
\newcommand{\tev}{\mbox{${\rm TeV}$}}
\newcommand{\gev}{\mbox{${\rm GeV}$}}
\newcommand{\mev}{\mbox{${\rm MeV}$}}

\newcommand{\QQbar}{\mbox{$\mathrm {Q\overline{Q}}$}}
\newcommand{\Dz}{\mbox{$\mathrm {D^0}$}}

\newcommand{\ppbar}{\mbox{$\mathrm {p\overline{p}}$}}
\newcommand{\ccbar}{\mbox{$\mathrm {c\overline{c}}$}}
\newcommand{\bbbar}{\mbox{$\mathrm {b\overline{b}}$}}
\newcommand{\sbbbar}{\mbox{$\scriptstyle\mathrm {b\overline{b}}$}}
\begin{document}
%%%
%%%  TITLEPAGE
%%%
\begin{titlepage}

\begin{center}
\raisebox{0.5cm}[0cm][0cm]{
\begin{tabular*}{\hsize}{@{\hspace*{-5mm}}ll@{\extracolsep{\fill}}r@{}}
\begin{minipage}[t]{3cm}
\vglue.5cm
\includegraphics[bb=226 315 386 476,width=2.4cm]{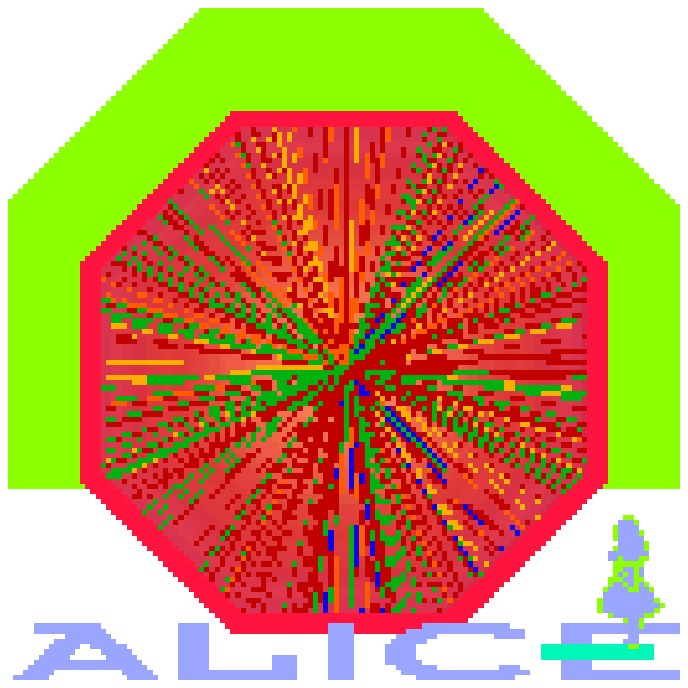}
\end{minipage}
&
\begin{minipage}[t]{5cm}
\vglue.5cm
%\institute{INSTITUTE}%%%%%Left header side to complete
%\instadre{Institute adress\\
%adress suite\\
%adress option\\
%adress option 2}%%%%%%%%End of the left header side
\end{minipage}
&
\begin{minipage}[t]{5cm}
\vglue.5cm
\note{Internal NOTE}%%%%%%%Right header side to complete
\vglue-.2cm
\refalice{ALICE Reference Number}
\notenum{ALICE-INT-2003-019 v.3}
\vglue-.2cm
%\refinstitute{Institute reference number}
%\Empty{[-]}
%\vglue-.1cm
\changedate{date of last change}
\vglue-.1cm
\notedate{18th November 2003}%%%%%%end of the right header side
\end{minipage}
\end{tabular*}
}
\end{center}

\vskip2.5cm

\vglue1cm
\title{\Large Charm and beauty production at LHC}

\begin{Authlist}

{N. Carrer}

\Instfoot{a1}{{\it CERN, 1211 Geneva, Switzerland}}
{A. Dainese}
\Instfoot{a1}{{\it Universit\`a degli Studi di Padova and INFN, 
              via Marzolo 8, 35131 Padova, Italy\\
              e-mail: andrea.dainese@pd.infn.it}} 

\end{Authlist}
\vglue2cm

\begin{abstract}
This note will be part of Chapter 6.5 of the ALICE Physics Performance
Report (PPR), {\sl ALICE Physics: Charm and Beauty}, where the capabilities 
of ALICE for the detection of open heavy flavour particles will be described.\\
We define here the present ALICE baseline for what concerns the heavy flavour 
production cross sections at LHC and the kinematical distributions 
of the heavy quark pairs. We start by qualitatively assessing the Bjorken $x$ 
regimes accessible with charm and beauty measurements at LHC with ALICE 
(Section~\ref{CHAP6_5:x1x2}). In Section~\ref{CHAP6_5:XsecNN}
we report the most recent results (and the uncertainties) 
of the next-to-leading order (NLO) pQCD calculations for the cross sections in 
pp collisions at LHC energies. These results are extrapolated to 
\mbox{Pb--Pb} collisions in Section~\ref{CHAP6_5:extrapolation2PbPb}
and to \mbox{p--Pb} collisions in Section~\ref{CHAP6_5:extrapolation2pPb},
taking into account nuclear shadowing and parton intrinsic transverse 
momentum broadening.
Heavy quark kinematics as given by the NLO pQCD calculation are reported in 
Section~\ref{CHAP6_5:kinematic}.  
We tuned the PYTHIA event generator in order to reproduce such results for 
what concerns the c and b quarks transverse momentum distributions
(Section~\ref{CHAP6_5:generators}). Finally, we report the yields and 
transverse momentum distributions for D and B mesons 
(Section~\ref{CHAP6_5:hadr}).
\end{abstract}

\vspace{7cm}
%\submitted{(.....................................)}
\end{titlepage}

\clearpage

\thispagestyle{empty}
~

\clearpage

\setcounter{page}{1}

\section{Accessible $x$ range with heavy quarks at LHC}
\label{CHAP6_5:x1x2}

In the inelastic collision of a proton (or, more generally, nucleon) with 
a particle, the Bjorken $x$ variable is defined 
as the fraction of the proton momentum carried by the parton that 
enters the hard scattering process. The distribution of $x$ for a given 
parton type (e.g. gluon, valence quark, sea quark) is called
Parton Distribution Function (PDF) and it gives the
probability to pick up a parton with momentum fraction $x$ from the proton.

The LHC will allow to probe the parton distribution functions of the nucleon 
and, 
in the case of \pA~and \AA~collisions, also their modifications
in the nucleus, down to unprecedented low values of $x$. 
In this paragraph we compare the regimes in $x$ corresponding to the 
production of a $\ccbar$ pair at SPS, RHIC and LHC energies and we
estimate the $x$ range that can be 
accessed with ALICE~\cite{alice} for what concerns heavy flavour production. 
This information is particularly valuable because
the charm and beauty production cross sections 
at the LHC are significantly affected by parton dynamics in the 
small-$x$ region, as we will see in the following sections. 
Therefore, the measurement of heavy flavour production may provide 
information on the nuclear parton densities. 

We can consider the simple case of the production of a heavy quark
pair, $Q\overline{Q}$, through the leading order\footnote{Leading order
(LO) is $\mathcal{O}(\alpha_s^2)$;
next-to-leading order (NLO) is $\mathcal{O}(\alpha_s^3)$.}  
gluon--gluon fusion process
$gg\to Q\overline{Q}$ in the collision 
of two ions $({\rm A}_1,{\rm Z}_1)$ and $({\rm A}_2,{\rm Z}_2)$. 
The $x$ range actually probed depends on the value of the centre-of-mass
(c.m.s.) energy
per nucleon pair $\sqrtsNN$, 
on the invariant mass\footnote{For two particles with 
four-momenta $(E_1,\vec{p}_1)$ and $(E_2,\vec{p}_2)$,
the invariant mass is defined as the modulus of the total four-momentum:
$M=\sqrt{(E_1+E_2)^2-(\vec{p}_1+\vec{p}_2)^2}$.}
$M_{Q\overline{Q}}$ of the $Q\overline{Q}$
pair produced in the hard scattering and on its rapidity 
$y_{Q\overline{Q}}$. If the parton intrinsic transverse momentum 
in the nucleon is neglected, the four-momenta of the two incoming gluons are 
$(x_1,0,0,x_1)\cdot ({\rm Z_1/A_1})\,\sqrt{s_{\rm pp}}/2$ and 
$(x_2,0,0,-x_2)\cdot ({\rm Z_2/A_2})\,\sqrt{s_{\rm pp}}/2$,
where $x_1$ and $x_2$ are the momentum fractions carried by the gluons, 
and $\sqrt{s_{\rm pp}}$ is the c.m.s. energy for pp 
collisions ($14~\tev$ at the LHC).
The square of the invariant mass of the $Q\overline{Q}$ pair is given by:
\begin{equation}
\label{eq:sx1x2M2}
  M^2_{Q\overline{Q}}=\hat{s}=x_1\,x_2\,s_{\scriptscriptstyle \rm NN}=x_1\,\frac{{\rm Z}_1}{{\rm A}_1}\,x_2\,\frac{{\rm Z}_2}{{\rm A}_2}\,s_{\rm pp};
\end{equation}
and its longitudinal rapidity in the laboratory is:
\begin{equation}
\label{eq:rapidityx1x2}
  y_{Q\overline{Q}} = \frac{1}{2}\ln \left[\frac{E+p_z}{E-p_z}\right] = \frac{1}{2}\ln\left[\frac{x_1}{x_2}\cdot\frac{{\rm Z}_1\,{\rm A}_2}{{\rm Z}_2\,{\rm A}_1}\right].
%\begin{array}{rl}
%  y_{Q\overline{Q}} & = \frac{1}{2}\ln \left[\frac{E+p_z}{E-p_z}\right] =\frac{1}{2}\ln \left[\frac{(x_1\frac{{\rm Z}_1}{{\rm A}_1}+x_2\frac{{\rm Z}_2}{{\rm A}_2})\cdot \sqrt{s_{\rm pp}}/2 +(x_1\frac{{\rm Z}_1}{{\rm A}_1}-x_2\frac{{\rm Z}_2}{{\rm A}_2})\cdot \sqrt{s_{\rm pp}}/2}{(x_1\frac{{\rm Z}_1}{{\rm A}_1}+x_2\frac{{\rm Z}_2}{{\rm A}_2})\cdot \sqrt{s_{\rm pp}}/2-(x_1\frac{{\rm Z}_1}{{\rm A}_1}-x_2\frac{{\rm Z}_2}{{\rm A}_2})\cdot \sqrt{s_{\rm pp}}/2}\right] \\
% & = \frac{1}{2}\ln\left[\frac{{\rm Z}_1\,{\rm A}_2}{{\rm Z}_2\,{\rm A}_1}\cdot\frac{x_1}{x_2}\right].
%\end{array}
\end{equation}

From these two relations we can derive the dependence of $x_1$ and $x_2$ on 
colliding system, $M_{Q\overline{Q}}$ and $y_{Q\overline{Q}}$:
\begin{equation}
  x_1 = \frac{{\rm A}_1}{{\rm Z}_1}\cdot\frac{M_{Q\overline{Q}}}{\sqrt{s_{\rm pp}}}\exp\left({+y_{Q\overline{Q}}}\right)~~~~~~~~~~~~~~~~ 
  x_2 = \frac{{\rm A}_2}{{\rm Z}_2}\cdot\frac{M_{Q\overline{Q}}}{\sqrt{s_{\rm pp}}}\exp\left({-y_{Q\overline{Q}}}\right); 
\end{equation}
which simplifies to
\begin{equation}
\label{eq:yx1x2}
  x_1 = \frac{M_{Q\overline{Q}}}{\sqrtsNN}\exp\left({+y_{Q\overline{Q}}}\right)~~~~~~~~~~~~~~~~ 
  x_2 = \frac{M_{Q\overline{Q}}}{\sqrtsNN}\exp\left({-y_{Q\overline{Q}}}\right)
\end{equation}
for a symmetric colliding system ($\rm A_1=A_2$, $\rm Z_1=Z_2$).

At central rapidities we have $x_1\simeq x_2$ and their magnitude
is determined by the ratio of the pair invariant mass to the c.m.s. energy.
For production at threshold 
($M_{\rm \scriptstyle {c\overline{c}}}=2\,m_{\rm c}\simeq 2.4$~GeV, 
$M_{\rm \scriptstyle {b\overline{b}}}=2\,m_{\rm b}\simeq 9$~GeV) 
we obtain what reported in Table~\ref{tab:xtable}. The $x$ regime 
relevant to charm production at the LHC ($\sim 10^{-4}$) is about 2 orders 
of magnitude lower than at RHIC and 3 orders of magnitude lower than at 
the SPS.

\begin{table}[!t]
  \caption{Bjorken $x$ values corresponding to charm and beauty production 
           at threshold at central rapidity.}
  \label{tab:xtable}
  \begin{center}
  \begin{tabular}{|c|cccc|}
\hline
Machine & SPS & RHIC & LHC & LHC \\
System & \PbPb & Au--Au & Pb--Pb & pp \\
$\sqrtsNN$ & 17 GeV & 200 GeV & 5.5 TeV & 14 TeV \\
\hline
\hline
$\ccbar$ & $x\simeq 10^{-1}$ & $x\simeq 10^{-2}$ & $x\simeq 4\cdot 10^{-4}$ &  $x\simeq 2\cdot 10^{-4}$ \\
$\bbbar$ & -- & -- & $x\simeq 2\cdot 10^{-3}$ & $x\simeq 6\cdot 10^{-4}$ \\
\hline
  \end{tabular}
  \end{center}
\end{table}

\begin{figure}[!t]
  \begin{center}
    \includegraphics[width=.78\textwidth]{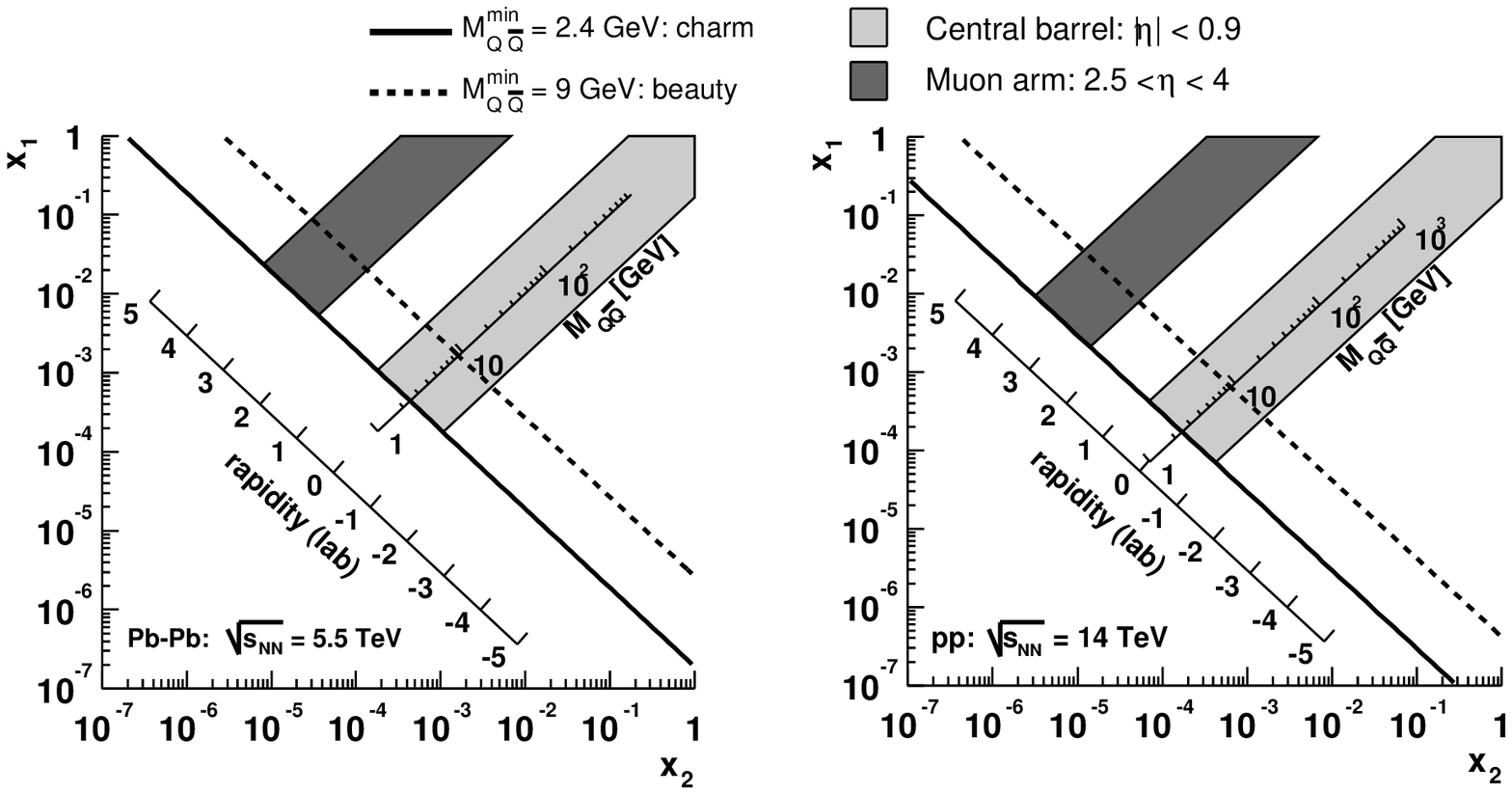}
  \caption{ALICE acceptance in the ($x_1$, $x_2$) plane for heavy flavours 
           in \mbox{Pb--Pb} (left) and in pp (right). The figure is explained 
           in detail in the text.}
  \label{fig:x1x2_AApp}
  \vglue0.2cm
    \includegraphics[width=.78\textwidth]{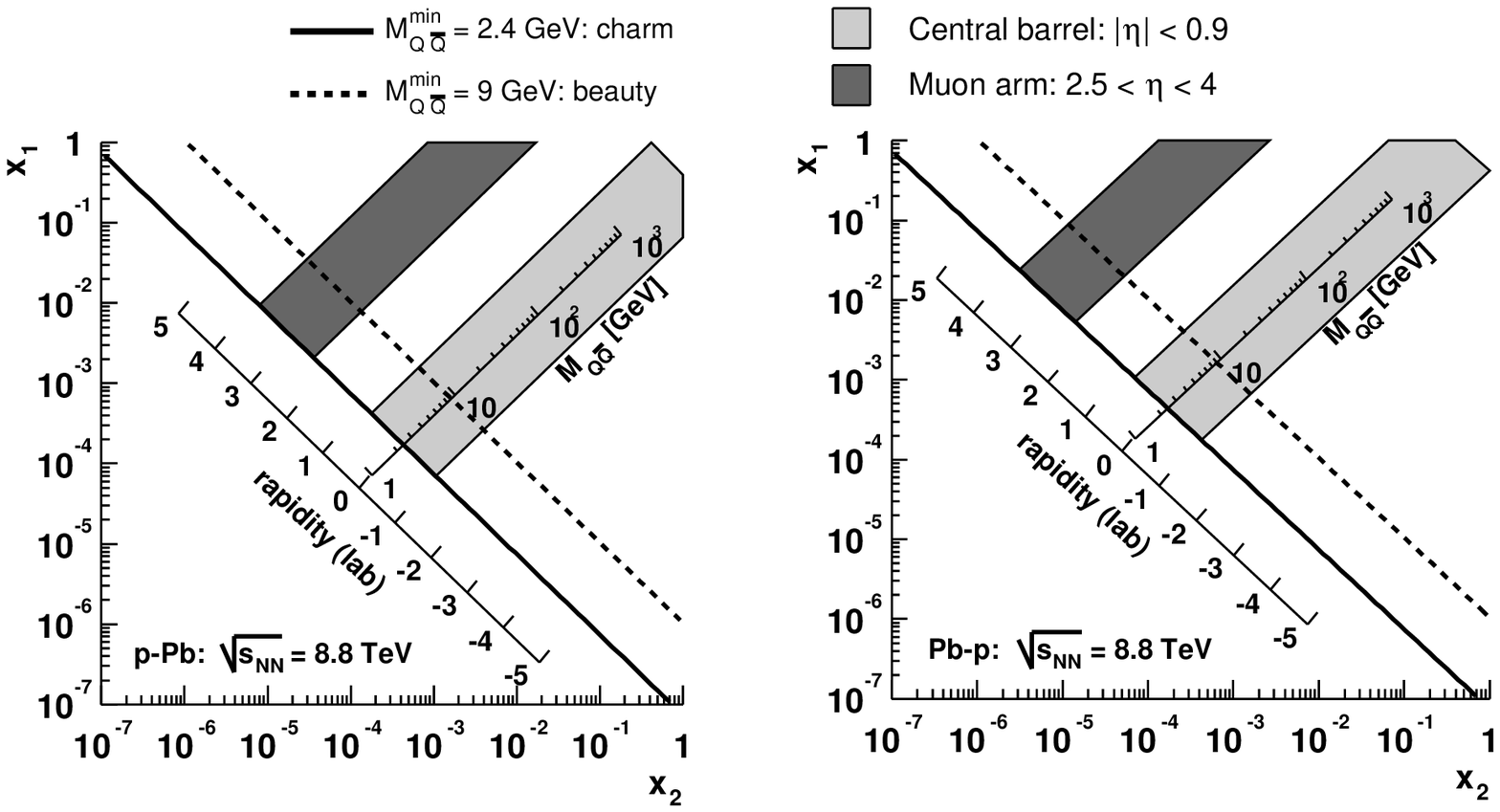}
  \caption{ALICE acceptance in the ($x_1$, $x_2$) plane for heavy flavours 
           in \mbox{p--Pb} (left) and in \mbox{Pb--p} (right).}
  \label{fig:x1x2_pAAp}
  \vglue0.2cm
    \includegraphics[width=.78\textwidth]{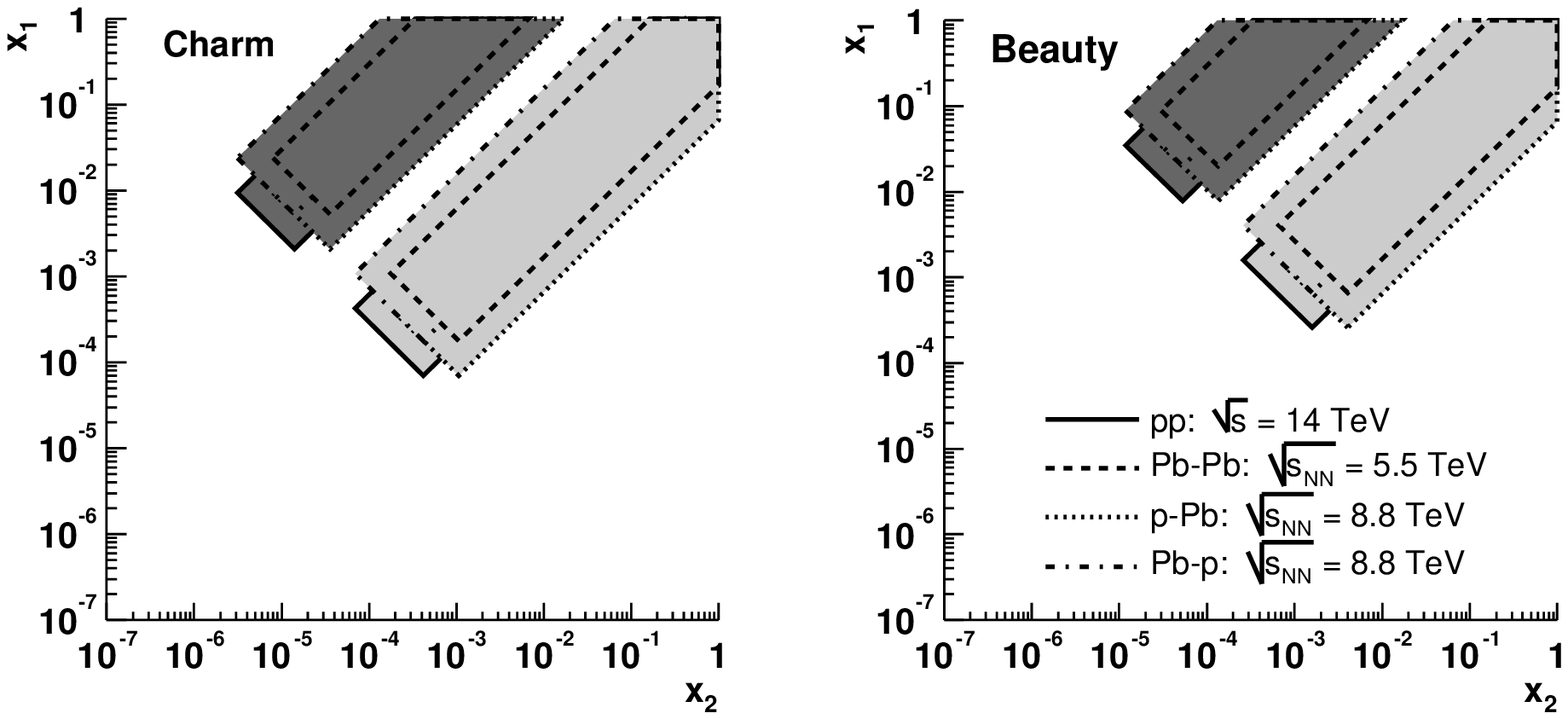}
  \caption{ALICE acceptance in the ($x_1$, $x_2$) plane for 
           charm (left) and beauty (right) in pp, \mbox{Pb--Pb}, 
           \mbox{p--Pb} and \mbox{Pb--p}.}
  \label{fig:x1x2_global}
  \end{center}
\end{figure}

Because of its lower mass, charm allows to probe lower $x$ values than beauty. 
The capability to measure charm and beauty particles in the forward 
rapidity region ($y\simeq 4$) would give access to $x$ regimes about 
2 orders of magnitude lower, down to $x\sim 10^{-6}$. 

In Fig.~\ref{fig:x1x2_AApp} we show the regions of the ($x_1$, $x_2$) plane 
covered for charm and beauty by the ALICE acceptance, 
in \mbox{Pb--Pb} at 5.5~TeV 
and in pp at 14~TeV. In this plane the points with constant invariant 
mass lie on hyperbolae 
($x_1=M^2_{Q\overline{Q}}/(x_2\,s_{\scriptscriptstyle \rm NN})$), 
straight lines in the log-log scale: 
we show those corresponding to the production of 
$\ccbar$ and $\bbbar$ pairs at the threshold; 
the points with constant rapidity lie
on straight lines ($x_1=x_2\exp(+2\,y_{Q\overline{Q}})$). The shadowed regions 
show the acceptance of the ALICE barrel, covering the 
pseudorapidity
range $|\eta|<0.9$, and 
of the muon arm, $2.5<\eta<4$.

In the case of asymmetric collisions, e.g. \mbox{p--Pb} and 
\mbox{Pb--p}\footnote{When we write \mbox{p--Pb}, 
we mean that the proton moves with $p_{\rm z}>0$; when we write \mbox{Pb--p}, 
we mean that the proton moves 
with $p_{\rm z}<0$.}, 
we have a rapidity shift: the centre of mass moves with a longitudinal 
rapidity
\begin{equation}
  y_{\rm c.m.} = \frac{1}{2}\ln\left(\frac{\rm Z_1 A_2}{\rm Z_2 A_1}\right),
\end{equation}
obtained from equation~(\ref{eq:rapidityx1x2}) for $x_1=x_2$. The rapidity 
window covered by the experiment is consequently shifted by
\begin{equation}
 \Delta y = y_{\rm lab.~system}-y_{\rm c.m.~system} = y_{\rm c.m.},
\end{equation}
corresponding to $+0.47$ ($-0.47$) for p--Pb (Pb--p) collisions.
Therefore, running with both \mbox{p--Pb} and \mbox{Pb--p} will allow 
to cover the largest interval in $x$. The c.m.s. energy in this case 
is 8.8~TeV.
Figure~\ref{fig:x1x2_pAAp} shows the acceptances for \mbox{p--Pb} and 
\mbox{Pb--p}, while in Fig.~\ref{fig:x1x2_global} the coverages 
in pp, \mbox{Pb--Pb}, \mbox{p--Pb} and \mbox{Pb--p} are 
compared for charm (left) and 
beauty (right). 

\begin{figure}[!t]
  \begin{center}
  \includegraphics[width=.6\textwidth]{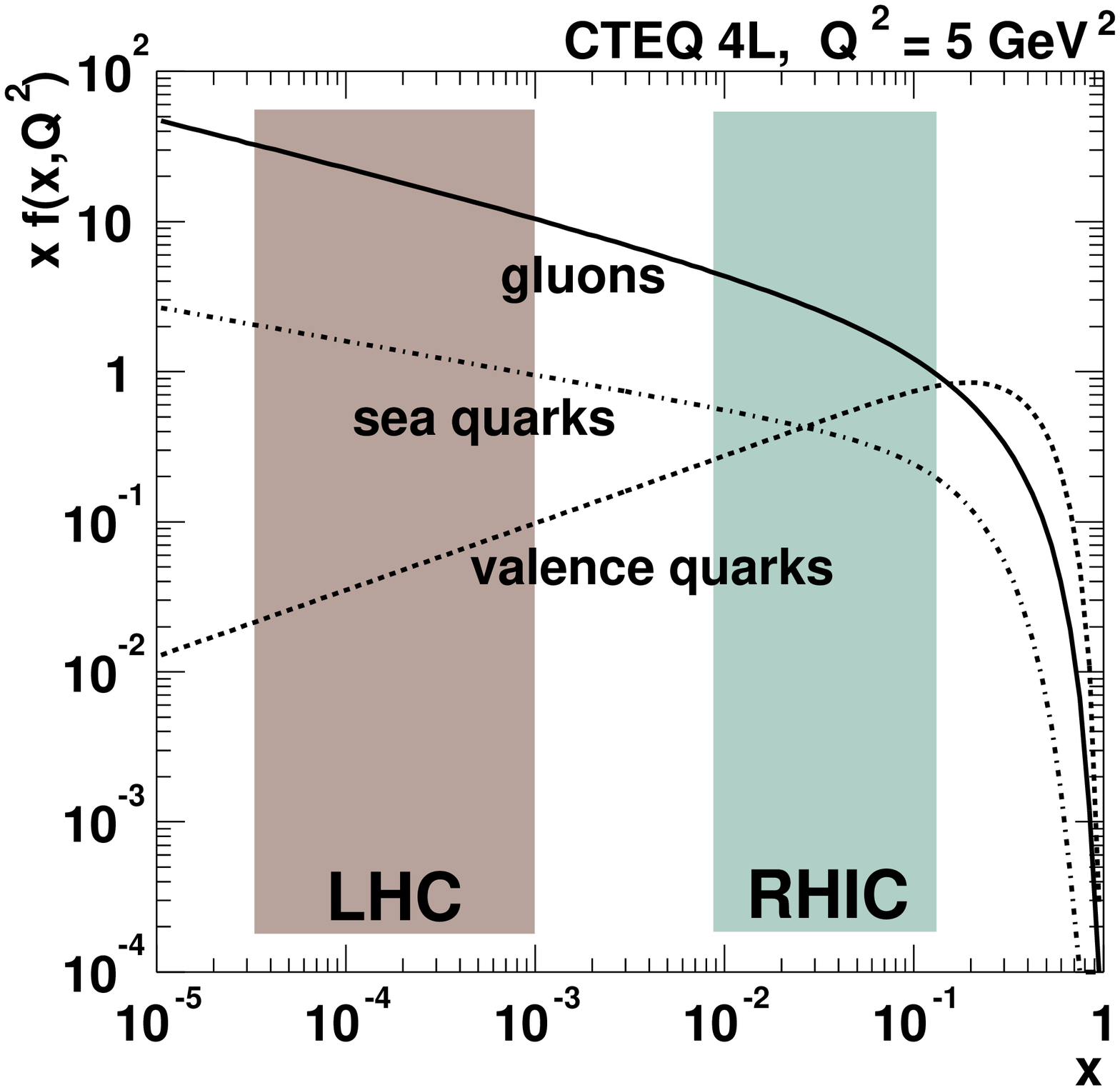}
  \caption{Parton distribution functions in the proton, in the CTEQ 4L 
           parameterization, for $Q^2=5~\gev^2$.} 
  \label{fig:pdf}
  \end{center}
\end{figure}
  
These figures are meant to give a first idea of the 
regimes accessible with ALICE; the simple relations for the leading order 
case were used, the ALICE rapidity acceptance cuts were applied to the 
rapidity of the $Q\overline{Q}$ pair, and not to that of the particles 
which are actually detected. In addition, no minimum $p_{\rm t}$ cuts 
were accounted for: such cuts will increase the minimum accessible value of 
$M_{Q\overline{Q}}$, thus increasing also the minimum accessible $x$. 
These approximations, however, are not too drastic, since 
there is a very strong correlation in rapidity between the 
initial $Q\overline{Q}$ pair and the heavy flavour particles it produces
and the minimum $p_{\rm t}$ 
cut will be quite low (lower than the mass of the hadron) for most of the 
channels studied with ALICE.

The parton distribution functions $x\,f(x,Q^2)$ in the proton, in the CTEQ 
4L parameterization~\cite{cteq4}, are shown in Fig.~\ref{fig:pdf}. 
$Q^2$ is the virtuality, 
or QCD scale (in the case of the leading order heavy flavour 
production considered in this paragraph, 
$Q^2=M^2_{Q\overline{Q}}=s\,x_1\,x_2$). In the figure the value 
$Q^2=5~\gev^2$, corresponding to $\ccbar$ production at threshold, is used.
The regions in $x$ covered, at central rapidities, 
at RHIC and LHC are indicated by the shaded areas.

\section{Cross sections in nucleon--nucleon collisions}
\label{CHAP6_5:XsecNN}

In this section we present the status of the cross section
calculations in nucleon--nucleon collisions and their comparison with
existing data up to a c.m.s. energy of $\simeq 65~\gev$. We then report the
results for LHC energies. The extrapolation to heavy ion
collisions is described in the next section.

\begin{figure}[!t]
  \begin{center}
    \includegraphics[width=0.67\textwidth]{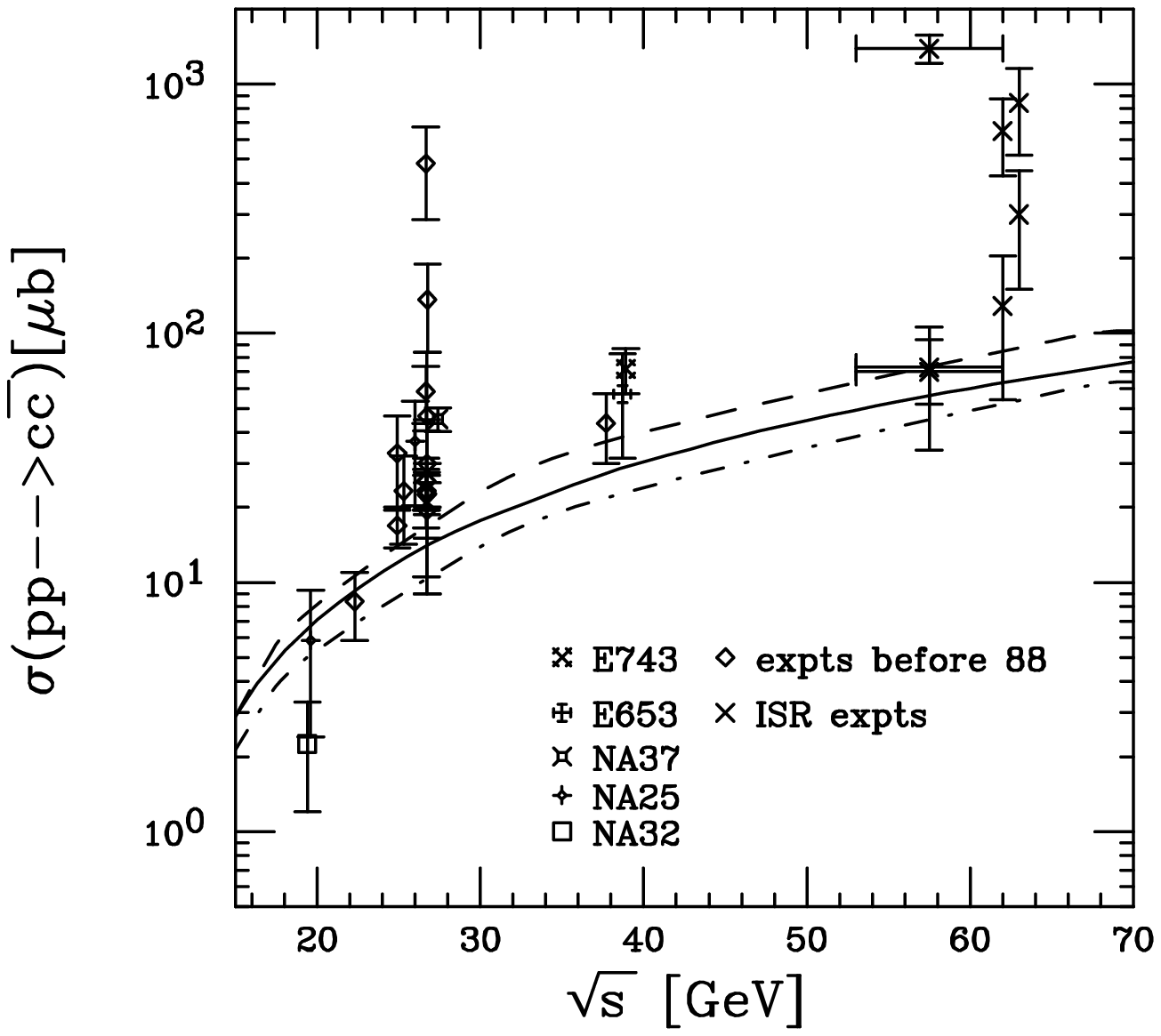}
    \caption{Total charm production cross section from pp and pA
    measurements compared to NLO calculations~\cite{vogtnew} with
    MRS D-' (solid), MRST HO (dashed) and MRST LO (dot-dashed) parton 
    distributions.}
    \label{fig:ramona1}
    \vglue0.5cm
    \includegraphics[angle=270,width=.49\textwidth]{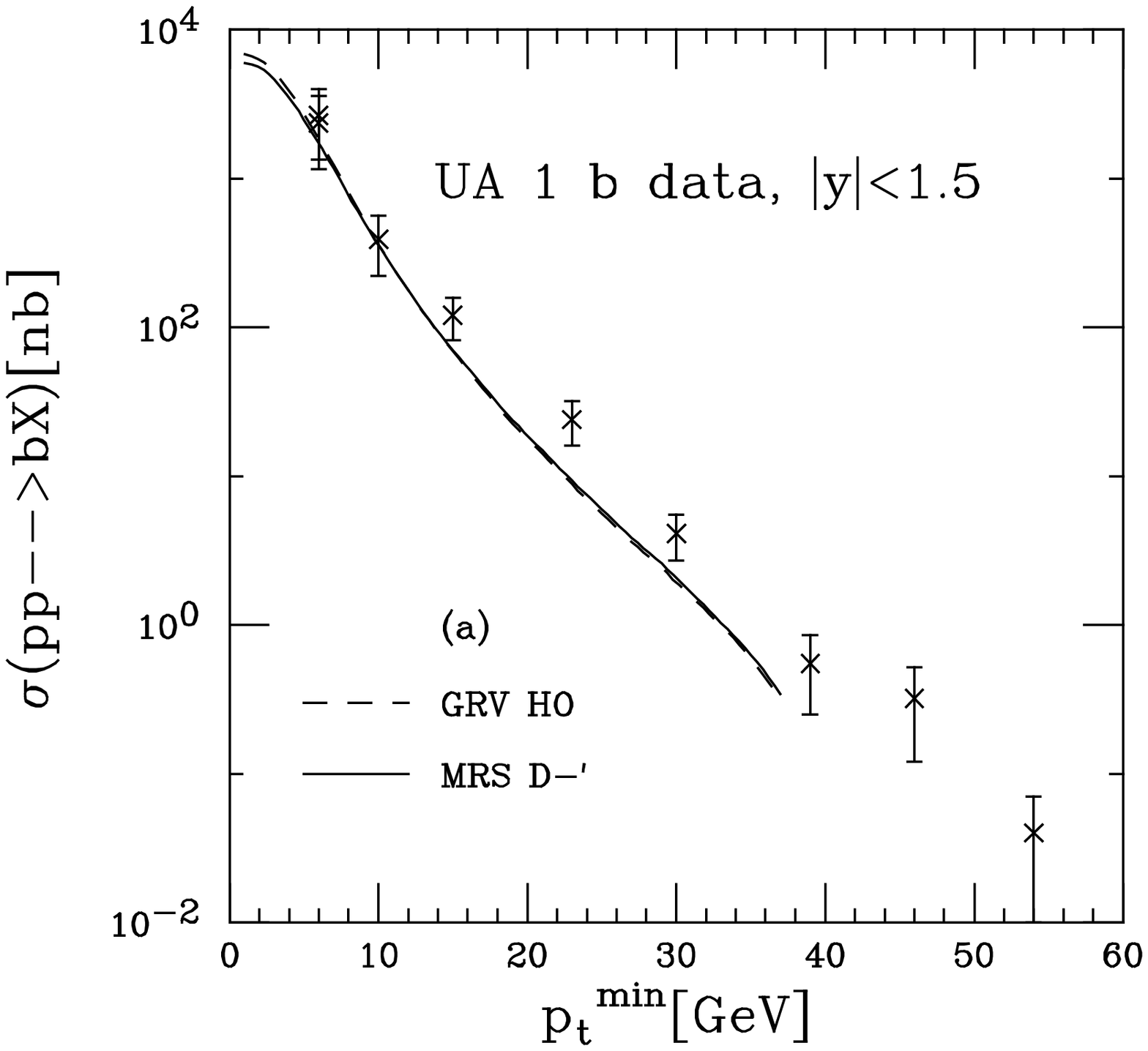}
    \includegraphics[angle=270,width=.49\textwidth]{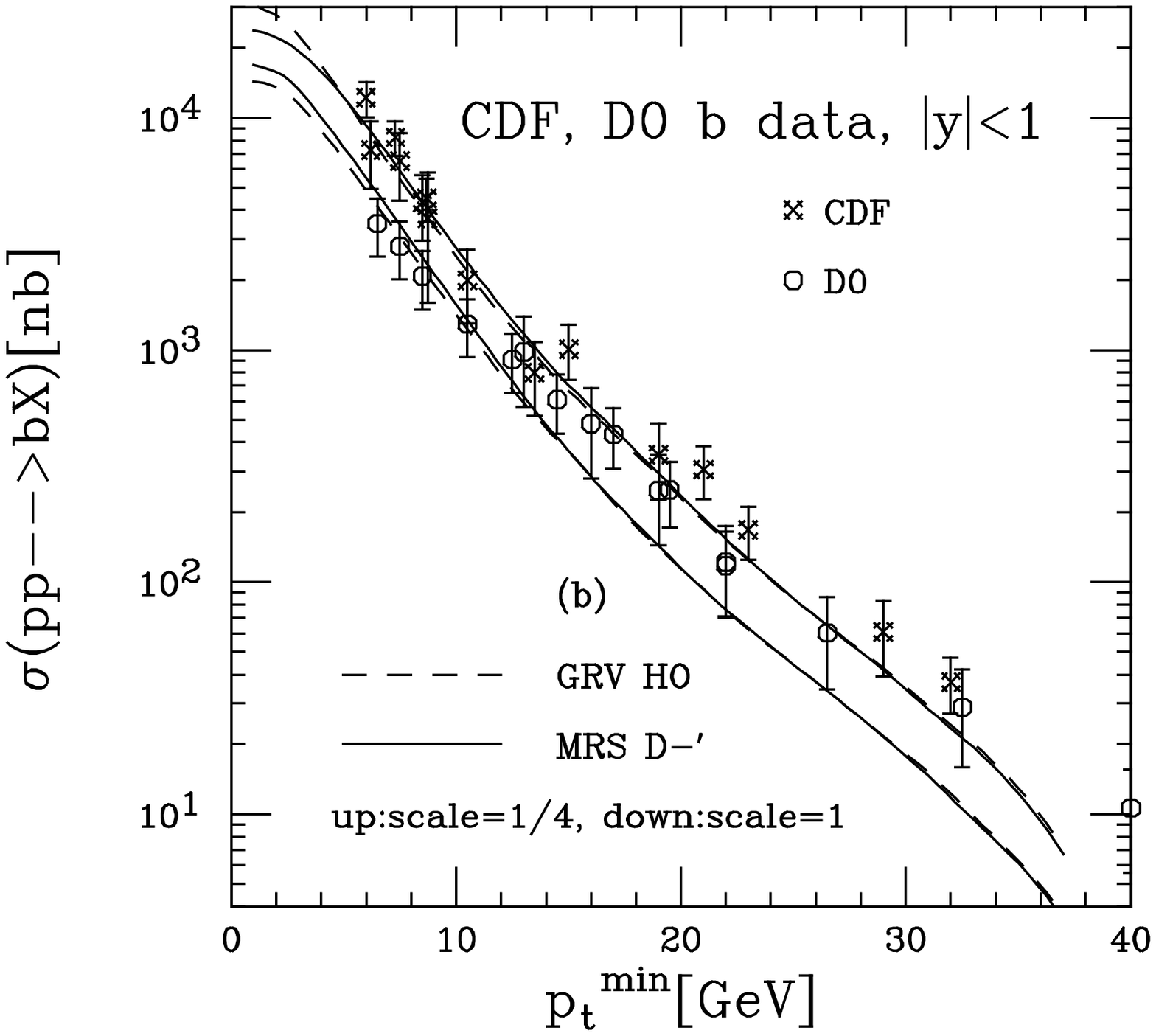}
    \caption{Comparison with b quark production cross section
    integrated over $\pt > \pt^{\rm min}$ from (a)
    UA1~\cite{UA1} and (b) CDF and D0~\cite{CDF}. The NLO calculations
    are with MRS D-' (solid) and GRV HO (dashed) parton
    distributions.}
    \label{fig:ramona2}
  \end{center}
\end{figure}

The existing data on total charm production cross section in pp and pA
collisions\footnote{The pA results were scaled according to the number 
of binary nucleon--nucleon collisions, in order to obtain the 
equivalent cross section in pp.} 
up to ISR energies are compared in Fig.~\ref{fig:ramona1}
with NLO calculations by R.~Vogt~\cite{vogtnew}. In
Fig.~\ref{fig:ramona2} a NLO calculation from Ref.~\cite{ramona94}
is compared to the data in \ppbar~collisions from UA1, CDF
and D0, for which the b quark production cross section integrated
for $\pt > \pt^{\rm min}$ is given. These measurements are taken in the
central rapidity region ($|y| < 1.5$ for UA1, $|y|<1$ for CDF and D0).
All the calculations have been performed using the following values
for the heavy quark masses ($m_{\rm c}$, $m_{\rm b}$) and for the factorization
and renormalization scales ($\mu_F$, $\mu_R$):
\begin{equation}
m_{\rm c} = 1.2~\gev
\,\,\,\,\,\,\,\,\,\,\,\,\,\,\,\,\,\,\,\,\,\,\,\,\,\,\,\,\,\,\,\,\,\,\,\,\,\,\,
\mu_F = \mu_R = 2\,\mu_0
\end{equation}
for charm, and
\begin{equation}
m_{\rm b} = 4.75~\gev
\,\,\,\,\,\,\,\,\,\,\,\,\,\,\,\,\,\,\,\,\,\,\,\,\,\,\,\,\,\,\,\,\,\,\,\,\,\,\,
\mu_F = \mu_R = \mu_0
\end{equation}
for beauty; 
$\mu_0\equiv\sqrt{(p_{\rm t,Q}^2+p_{\rm t,\overline{Q}}^2)/2+m_Q^2}$ 
is approximately equal to the transverse mass of the produced heavy quarks.

For both charm and beauty the theory describes the present data reasonably
well.

The results for LHC energies ($\sqrt{s}=$ 5.5, 8.8 and 14~$\tev$) 
are reported in Table~\ref{tab:lhcXsec}. 
These values are obtained using the
NLO pQCD calculation implemented in the program
by M.~Mangano, P.~Nason and G.~Ridolfi~\cite{MNRcode} (HVQMNR) and two sets of
parton distribution functions, MRST HO~\cite{mrst} and CTEQ 5M1~\cite{cteq5}, 
which include the small-$x$ HERA results. 
The difference due to the choice of the
parton distribution functions is relatively small ($\sim 20$-$25\%$ at
5.5~$\tev$, slightly lower at 14~$\tev$). We chose to use as a baseline 
the average, also reported in the table, of the values obtained 
with these two sets of PDF.

\begin{table}[!h]
\caption{NLO calculation~\cite{MNRcode} for the total \ccbar~
  and \bbbar~
  cross sections in pp collisions at 5.5, 8.8 and 14~$\tev$, using the MRST HO
  and CTEQ 5M1 parton distribution functions.}
\label{tab:lhcXsec}
\begin{center}
\begin{tabular}{|c|ccc|ccc|}
\hline
 & & $\sigma_{{\rm pp}}^{\scriptstyle{\rm c\overline{c}}} [{\rm mb}]$ & & & $\sigma_{{\rm pp}}^{\scriptstyle{\rm b\overline{b}}} [{\rm mb}]$ & \\
\hline
$\sqrt{s}$ & $5.5~\tev$ & $8.8~\tev$ & $14~\tev$ &  $5.5~\tev$ & $8.8~\tev$ & $14~\tev$ \\
\hline
\hline
MRST HO & 5.9 & 8.4 & 10.3 & 0.19 & 0.28 & 0.46 \\
CTEQ 5M1 & 7.4 & 9.6 & 12.1 & 0.22 & 0.31 & 0.55 \\
\hline
Average & 6.6 & 9.0 & 11.2 & 0.21 & 0.30 & 0.51 \\  
\hline
\end{tabular}
\end{center}
\end{table}

\begin{table}[!t]
\caption{Charm and beauty total cross sections at NLO with different
  choices of the parameters $m_{\rm c}$ ($m_{\rm b}$), $\mu_F$ and
  $\mu_R$~\cite{yelrepHardProbes}. In the last column the ratio 
  of the cross sections at 5.5~TeV and at 14~TeV is reported.}
\label{tab:manganoXsec}
\begin{center}
\begin{tabular}{|cc|c|c|c|}
\hline
& parameters & 5.5~$\tev$ & 14~$\tev$ & ratio 5.5~TeV/14~TeV \\
\hline
\hline
& $m_{\rm c} = 1.5~\gev$ & & &\\
& $\mu_R=2\mu_0 \,\,\,\,\,\, \mu_F=2\mu_0$ &
\raisebox{1.5ex}[0cm][0cm]{3.7} &
\raisebox{1.5ex}[0cm][0cm]{7.3} &
\raisebox{1.5ex}[0cm][0cm]{0.51} \\
\cline{2-5}
& $m_{\rm c} = 1.2$~GeV & & &\\
& $\mu_R=\mu_0 \,\,\,\,\,\, \mu_F=2\mu_0$ &
\raisebox{1.5ex}[0cm][0cm]{9.2} &
\raisebox{1.5ex}[0cm][0cm]{16.7} &
\raisebox{1.5ex}[0cm][0cm]{0.55} \\
\cline{2-5}
& $m_{\rm c} = 1.5$~GeV & & &\\
& $\mu_R=\mu_0 \,\,\,\,\,\, \mu_F=2\mu_0$ &
\raisebox{1.5ex}[0cm][0cm]{5.4} &
\raisebox{1.5ex}[0cm][0cm]{10.4} &
\raisebox{1.5ex}[0cm][0cm]{0.52} \\
\cline{2-5}
& $m_{\rm c} = 1.8~\gev$ & & &\\
\raisebox{10.5ex}[0cm][0cm]{$\sigma_{{\rm pp}}^{\scriptstyle{\rm c\overline{c}}} [{\rm mb}]$} &
$\mu_R=\mu_0 \,\,\,\,\,\, \mu_F=2\mu_0$ &
\raisebox{1.5ex}[0cm][0cm]{3.4} &
\raisebox{1.5ex}[0cm][0cm]{6.8} &
\raisebox{1.5ex}[0cm][0cm]{0.50} \\
\cline{2-5}
\hline
\hline
& $m_{\rm b} = 4.5~\gev$ & & &\\
& $\mu_R=\mu_0 \,\,\,\,\,\, \mu_F=\mu_0$ &
\raisebox{1.5ex}[0cm][0cm]{0.20} &
\raisebox{1.5ex}[0cm][0cm]{0.51} &
\raisebox{1.5ex}[0cm][0cm]{0.39} \\
\cline{2-5}
& $m_{\rm b} = 4.75~\gev$ & & &\\
& $\mu_R=\mu_0 \,\,\,\,\,\, \mu_F=\mu_0$ &
\raisebox{1.5ex}[0cm][0cm]{0.17} &
\raisebox{1.5ex}[0cm][0cm]{0.43} &
\raisebox{1.5ex}[0cm][0cm]{0.40} \\
\cline{2-5}
& $m_{\rm b} = 5~\gev$ & & &\\
& $\mu_R=\mu_0 \,\,\,\,\,\, \mu_F=\mu_0$ &
\raisebox{1.5ex}[0cm][0cm]{0.15} &
\raisebox{1.5ex}[0cm][0cm]{0.37} &
\raisebox{1.5ex}[0cm][0cm]{0.41} \\
\cline{2-5}
& $m_{\rm b} = 4.75~\gev$ & & \\
& $\mu_R=0.5\mu_0 \,\,\,\,\,\, \mu_F=2\mu_0$ &
\raisebox{1.5ex}[0cm][0cm]{0.26} &
\raisebox{1.5ex}[0cm][0cm]{0.66} &
\raisebox{1.5ex}[0cm][0cm]{0.39} \\
\cline{2-5}
& $m_{\rm b} = 4.75~\gev$ & & &\\
\raisebox{13.5ex}[0cm][0cm]{$\sigma_{{\rm pp}}^{\sbbbar} [{\rm mb}]$} &
$\mu_R=2\mu_0 \,\,\,\,\,\, \mu_F=0.5\mu_0$ &
\raisebox{1.5ex}[0cm][0cm]{0.088} &
\raisebox{1.5ex}[0cm][0cm]{0.20} &
\raisebox{1.5ex}[0cm][0cm]{0.44} \\
\hline
\end{tabular}
\end{center}
\end{table}

The dependence on the PDF set represents only a
part of the error on the theoretical estimate. An evaluation of the
theoretical uncertainties was done by M.~Mangano by varying the
$m_{\rm c}$ ($m_{\rm b}$), $\mu_F$ and $\mu_R$ parameters and is reported in
Table~\ref{tab:manganoXsec}~\cite{yelrepHardProbes}. This table shows that, at
LHC energies, the
theoretical uncertainties span a factor $\sim 2$-$3$ in the total
production cross section of both charm and beauty quarks.
In the last column of the table we report the ratio of the cross section 
at 5.5~TeV to that at 14~TeV. Despite the large spread of the 
absolute values, the ratio is much less dependent on the choice of the 
parameters; its value is $\simeq 0.52$ for charm and $\simeq 0.41$ for beauty. 
In Ref.~\cite{yelrepHardProbes} it is shown that also the ratios of the 
$\pt$-differential cross sections are rather independent of the parameters
choice.
This indicates that pQCD can be used to compare the cross 
sections measured in \PbPb~collisions at $\sqrtsNN=5.5~\tev$ to 
those measured in \pPb~at 
$\sqrtsNN=8.8~\tev$ and in pp at $\sqrt{s}=14~\tev$.

\subsubsection*{Yields in \pp~collisions at $\sqrt{s}=14~\tev$}

Using a \pp ~inelastic cross section 
$\sigma^{\rm inel}_{{\rm pp}}=70~{\rm mb}$ 
at 14~$\tev$~\cite{pprCh2} and the average heavy flavour cross sections
in the last row of Table~\ref{tab:lhcXsec}, 
we calculate the yields for the production of $Q\overline{Q}$ pairs as:
\begin{equation}
  N_{{\rm pp}}^{\scriptstyle Q\overline{Q}}=\sigma_{{\rm pp}}^{\scriptstyle Q\overline{Q}}\bigg/\sigma^{\rm inel}_{{\rm pp}}.
\end{equation}
We obtain 0.16~\ccbar~pairs and 0.0072~\bbbar~pairs per event.

\section{Extrapolation to heavy ion collisions}
\label{CHAP6_5:extrapolation}

In this section we derive the extrapolation of the cross sections and yields 
to central \PbPb~collisions first, and then to \pPb~collisions.
We also point out the different weight of the nuclear shadowing effect 
in the two cases, for charm and beauty production.

\subsection{Nucleus--nucleus collisions}
\label{CHAP6_5:extrapolation2PbPb}

If no nuclear effects are taken into account, a \AA~collision 
can be considered, for what concerns the hard cross section, 
as a superposition of {\sl independent}
\NN~collisions. Thus, the cross section for hard processes in heavy ion 
collisions can be calculated using a simple geometrical extrapolation 
from pp collisions, i.e. assuming that the hard cross section scales 
from pp to \AA~collisions proportionally to the number of 
inelastic \NN~collisions (binary scaling).

Nuclear effects ---such as nuclear shadowing, broadening of the
parton intrinsic transverse momentum ($k_{\rm t}$) in the nucleon, 
in-medium parton energy loss,
as well as possible enhancements due to thermal
production in the medium--- can modify this geometrical scaling from pp to
nucleus--nucleus collisions. Such effects are, indeed, what we want to measure.
We chose to include in the simulation only the nuclear shadowing
and the broadening of the intrinsic $k_{\rm t}$, 
since they are well established
effects. The first effect modifies the 
total hard cross section, while the broadening of the intrinsic $k_{\rm t}$ 
affects only the kinematic distributions of the produced heavy quarks. 
Nuclear shadowing can be accounted for by recalculating the 
hard cross section in elementary \NN~collisions with nuclear-modified parton 
distribution functions and extrapolating to the \AA~case.

The extrapolation, based on the Glauber model~\cite{glauber,cywong}, 
is derived for the collision of two generic nuclei 
with mass numbers A and B, and numerical examples are given 
for the specific case of \PbPb~reactions at $\sqrtsNN=5.5~\tev$.

We are interested in the cross section for a sample of events in a given 
centrality range, defined by the trigger settings. 
The centrality selection can be assumed to correspond to a cut on the 
impact parameter $b$ of the collision: $0\leq b < b_c$. 
The sample of events defined by this cut contains a fraction of the 
total number of inelastic collisions, i.e. of the total inelastic 
cross section, given by 
\begin{equation}
F(b_c)=\int_0^{b_c}{\rm d}b\,\frac{{\rm d}\sigma^{\rm inel}_{\rm AB}}{{\rm d}b}\bigg/\int_0^{\infty}{\rm d}b\,\frac{{\rm d}\sigma^{\rm inel}_{\rm AB}}{{\rm d}b}.
\end{equation}
The definition of the centrality in terms of fraction of the inelastic 
cross section is more appropriate, since the cross section is directly 
measured, while the impact parameter estimation depends on the model used to 
describe the geometry of the collision.

In the following, we consider two options for the centrality selection, 
corresponding to 5\% and 10\% of the total inelastic cross section. 
The values of $b_c$ that give these selections are $3.5~\fm$ and $5~\fm$,
respectively.

\begin{figure}
  \begin{center}
  \includegraphics[width=0.49\textwidth]{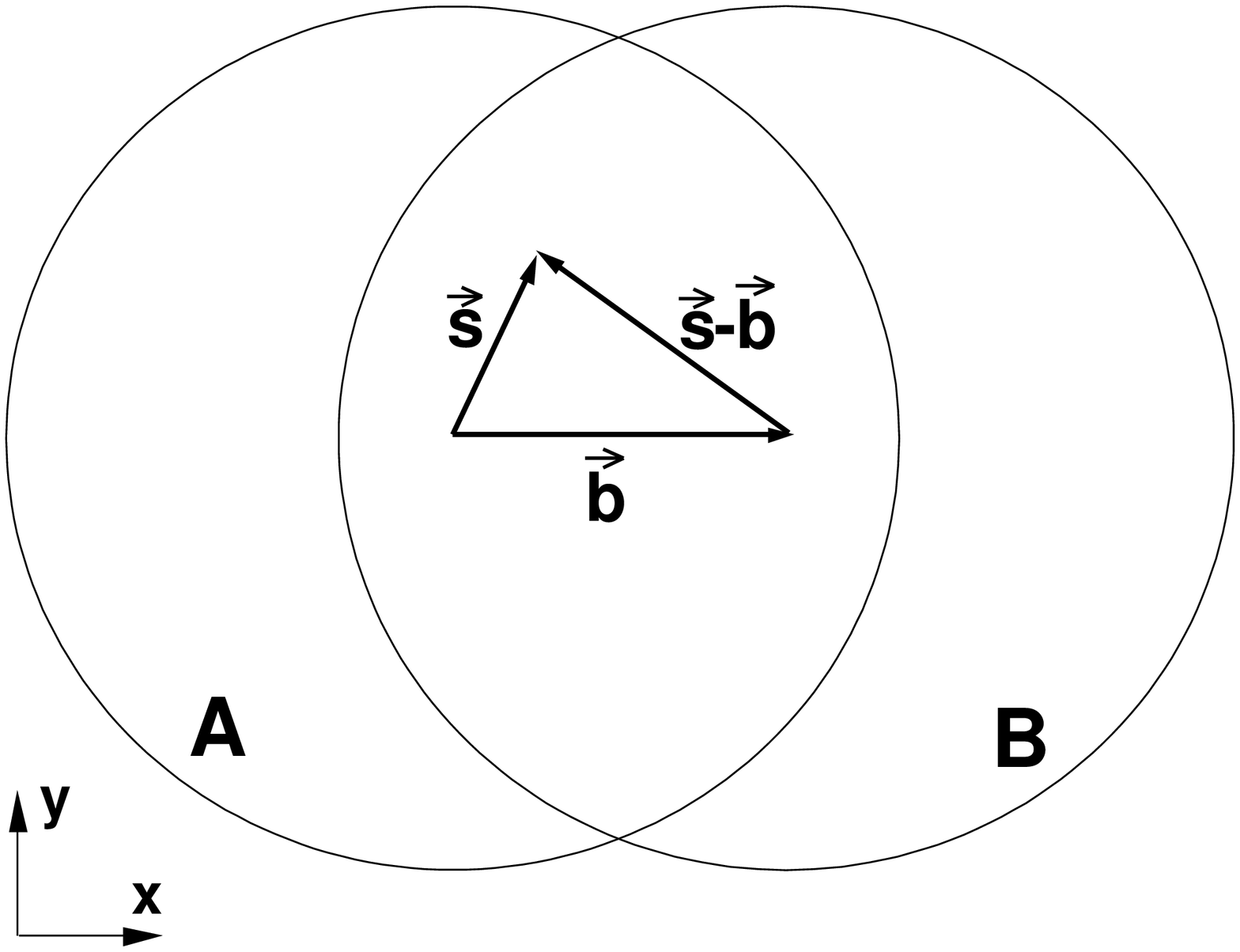}
  \includegraphics[width=0.49\textwidth]{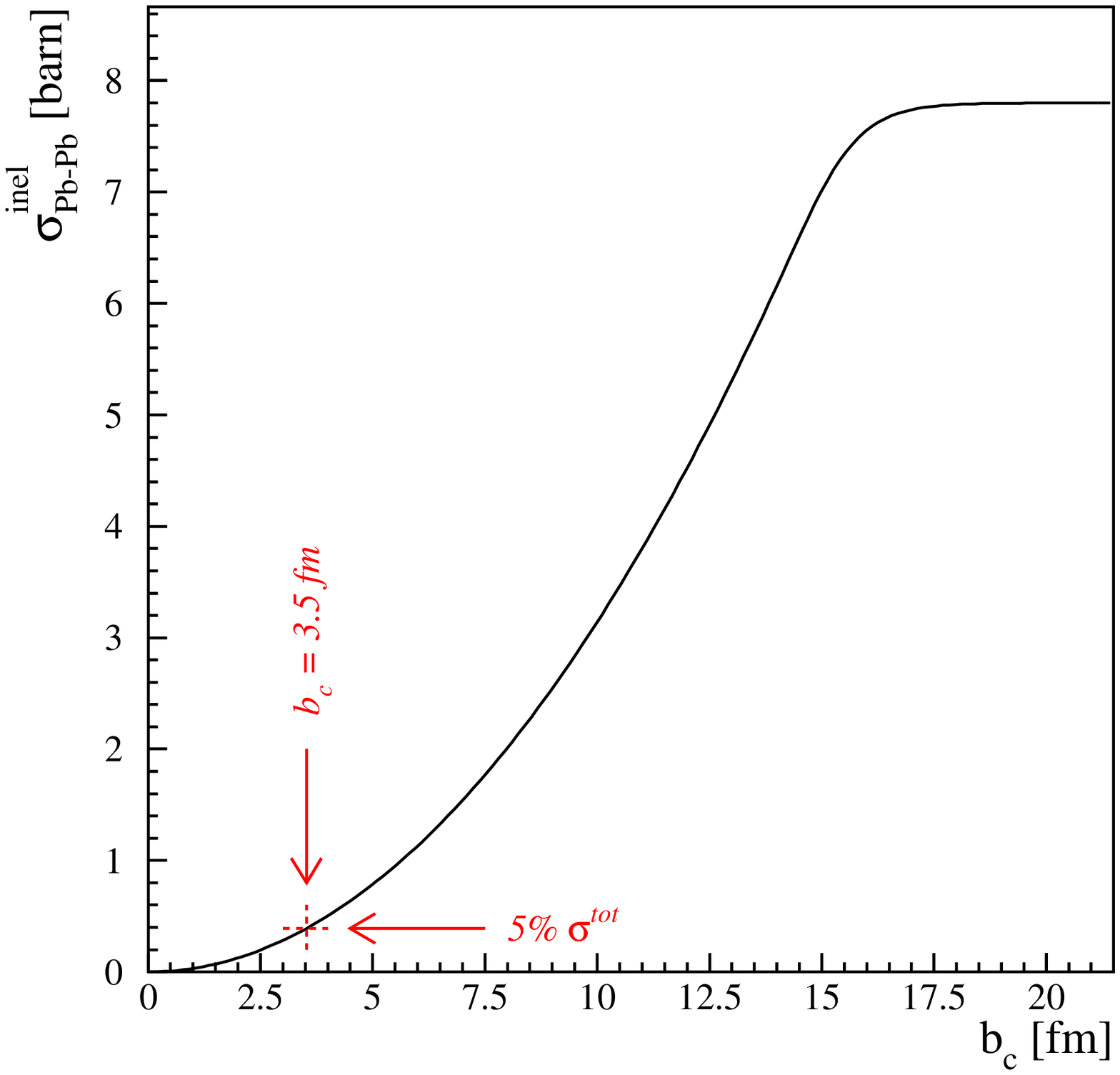}
  \caption{Collision geometry in the plane transverse to the beam line (left).
           Inelastic \PbPb~cross section as a function of the impact 
           parameter cut $b<b_c$ (right); for clearness, here and in 
           Fig.~\ref{fig:fhardANDrbc}, only
           the value corresponding to 5\% of the 
           total inelastic cross section is explicitly indicated.}
  \label{fig:totinel}
  \end{center}
\end{figure}

The inelastic cross section corresponding to a given centrality
selection is found integrating the interaction probability up to
impact parameter $b_c$:
\begin{equation}
  \sigma_{{\rm AB}}^\mathrm{inel}(b_c) = \int_0^{b_c}{\rm d}b\,\frac{{\rm d}\sigma^{\rm inel}_{\rm AB}}{{\rm d}b} = 2\pi \int_0^{b_c} b\,{\rm d}b\,\, 
  \left\{1 - [1 - \sigma_{\scriptscriptstyle{{\rm NN}}} T_{{\rm AB}}(b)]^{{\rm AB}} \right\}
  \label{eq:totinel}
\end{equation}
where the value $\sigma_{\scriptscriptstyle{{\rm NN}}} = 60~{\rm mb}$ 
was used as the
nucleon--nucleon inelastic cross section at 5.5 TeV~\cite{pdg}, 
and the total thickness function $T_{{\rm AB}}$
\begin{equation}
  T_{{\rm AB}}(b) = \int {\rm d}^2s\, T_{\rm A}(\vec{s})\,T_{\rm B}(\vec{s}-\vec{b})
\end{equation}
(vectors defined as in Fig.~\ref{fig:totinel}, left) 
is expressed in terms of the thickness function of the nucleus
$T_i(\vec{s}) = \int {\rm d}z\, \rho_i(z,\vec{s})$ for $i={\rm A,\,B}$, 
where $\rho_i$ is the Wood-Saxon nuclear density profile~\cite{woodsaxon} 
---the thickness function is 
normalized to unity: $\int {\rm d^2}s\,T_i(\vec{s})=1$. 
In Fig.~\ref{fig:totinel} (right) the
inelastic cross section (\ref{eq:totinel}) is shown as a function of $b_c$. 

The average number of inelastic collisions for a given impact 
parameter $b$ is:
\begin{equation}
\label{eq:collisions}
  \sigma_{\scriptscriptstyle{{\rm NN}}}\cdot {\rm AB}\,\,T_{{\rm AB}}(b).
\end{equation}
By replacing the inelastic \NN~cross section 
$\sigma_{\scriptscriptstyle{{\rm NN}}}$ with the 
elementary cross section for a given hard process 
$\sigma_{\rm pp}^{\rm hard}$, we obtain 
the average number of inelastic collisions that yield the considered hard 
process:
\begin{equation}
\label{eq:hardcollisions}
  \sigma_{\rm pp}^{\rm hard}\cdot {\rm AB}\,\,T_{{\rm AB}}(b), 
\end{equation}
and the cross section for hard processes for $0\leq b<b_c$:
\begin{equation}
\label{eq:sigmaABhard}
  \sigma_{{\rm AB}}^\mathrm{hard}(b_c) = \sigma_{{\rm pp}}^\mathrm{hard}\cdot 2\pi \,\int_0^{b_c} b\,{\rm d}b\,\,{\rm AB}\,\,
  T_{{\rm AB}}(b).
\end{equation}
For minimum-bias collisions ($b_c=+\infty$), we have:
\begin{equation}
\label{eq:proptoAB}
  \sigma_{{\rm AB}}^\mathrm{hard} = \sigma_{{\rm pp}}^\mathrm{hard}\,{\rm AB}.
\end{equation}

\begin{figure}
  \begin{center}
    \includegraphics[width=0.49\textwidth]{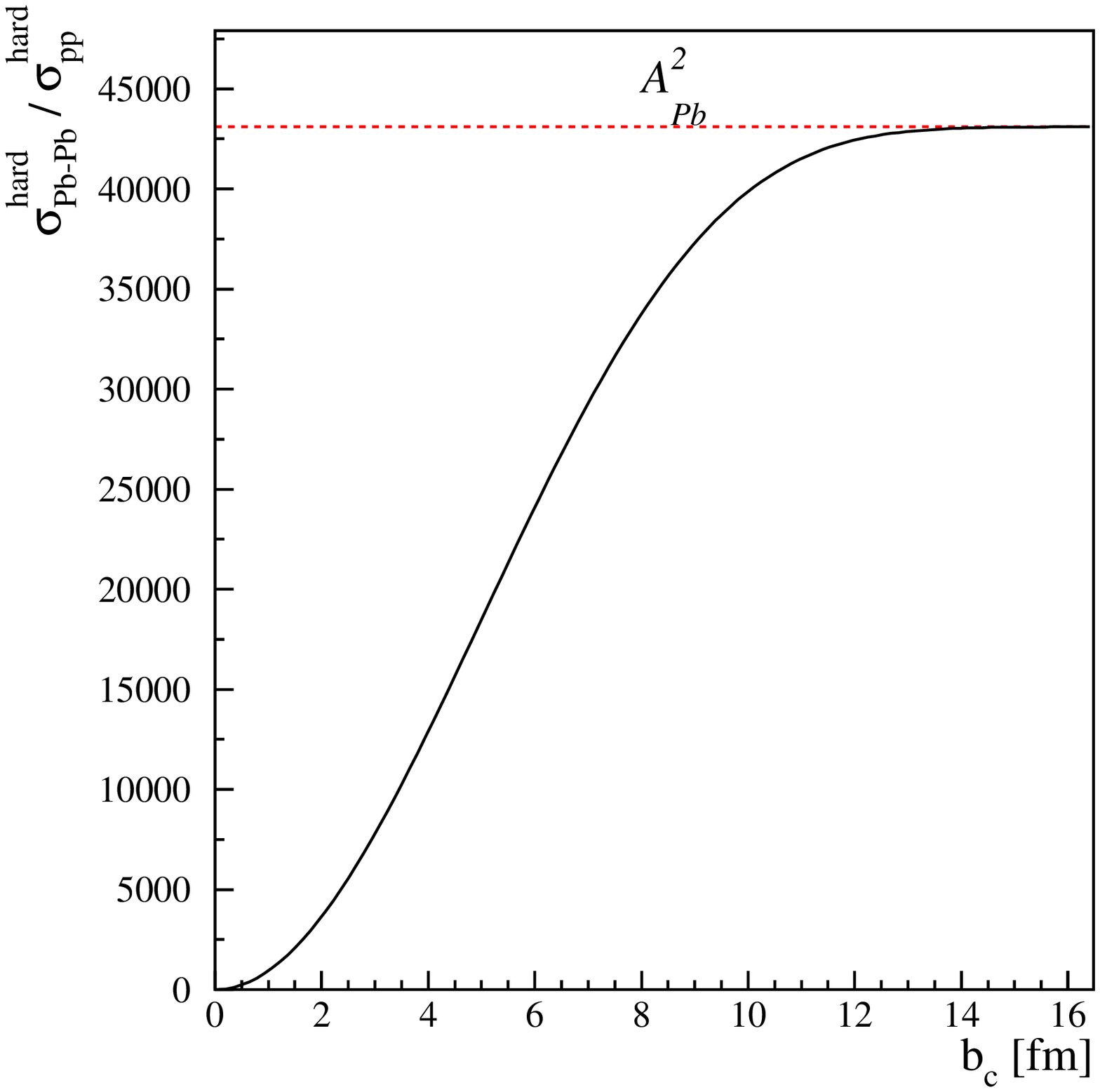}
    \includegraphics[width=0.49\textwidth]{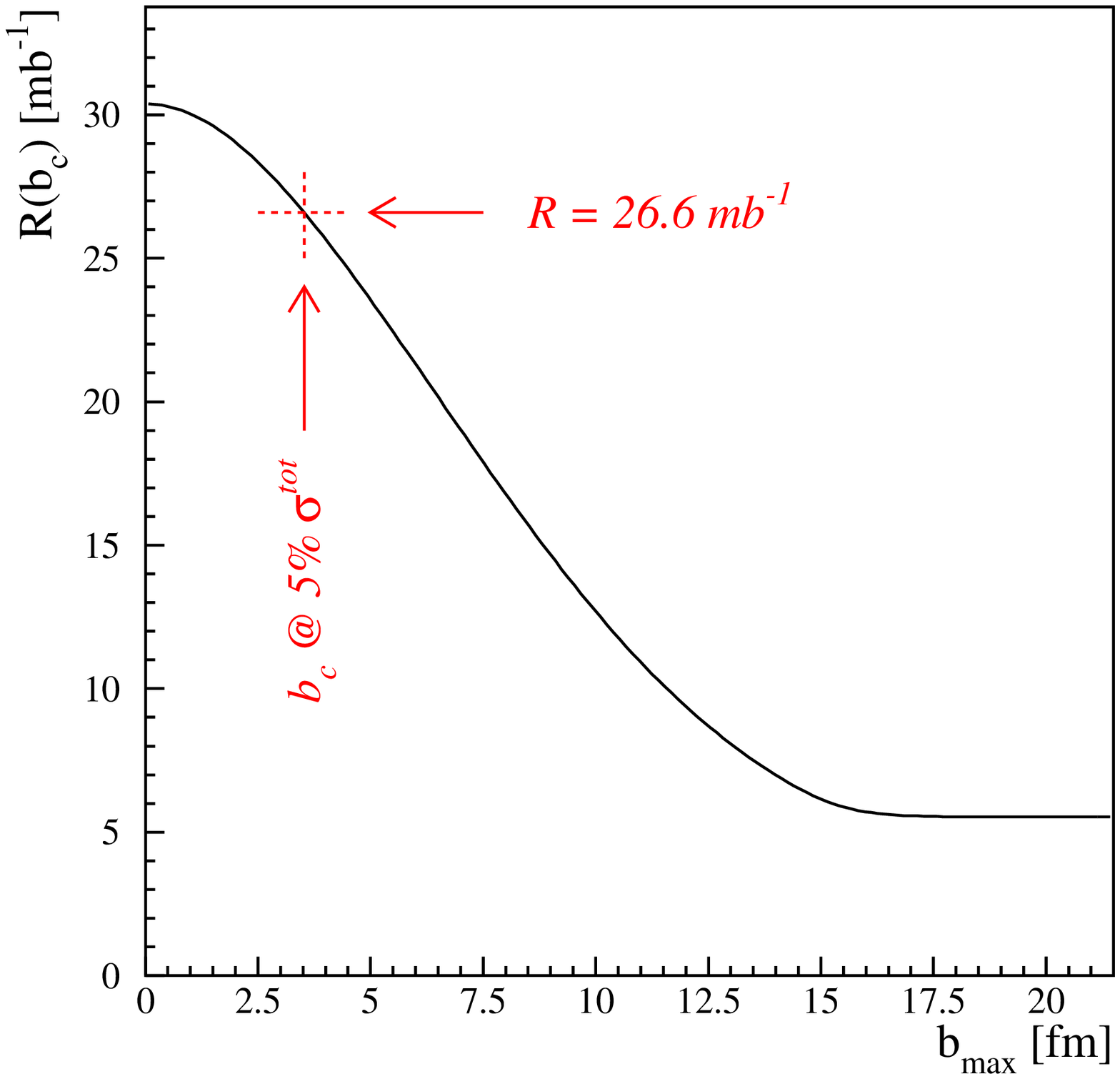}
  \caption{Cross section for a hard process in \mbox{Pb--Pb} collisions
    relative to the one in nucleon--nucleon collisions as a function of
    the impact parameter cut $b<b_c$ (left). Yield of the hard process in 
    \mbox{Pb--Pb} collisions
    relative to the cross section in nucleon--nucleon collisions as a
    function of the impact parameter cut $b<b_c$ (right).}
  \label{fig:fhardANDrbc}
  \end{center}
\end{figure}

The ratio of the hard cross section in nucleus--nucleus collisions, with a
centrality cut $b<b_c$, relative to the cross section in nucleon--nucleon
interactions is (see Fig.~\ref{fig:fhardANDrbc}, left):
\begin{equation}
  f^\mathrm{hard}(b_c) =
  \frac{\sigma_{{\rm AB}}^\mathrm{hard}(b_c)}{\sigma_{{\rm pp}}^\mathrm{hard}} =
  2\pi \int_0^{b_c} b\,{\rm d}b\,\, {\rm AB}\,\, T_{{\rm AB}}(b).
\end{equation}

The number (yield) of hard processes per triggered event is:
\begin{equation}
  N_{{\rm AB}}^\mathrm{hard}(b_c) = \frac{\sigma_{{\rm AB}}^\mathrm{hard}(b_c)}{\sigma_{{\rm AB}}^\mathrm{inel}(b_c)} = \mathcal{R}(b_c) \cdot \sigma_{{\rm pp}}^\mathrm{hard}
\end{equation}
where (Fig.~\ref{fig:fhardANDrbc}, right)
\begin{equation}
\label{eq:rbc}
  \mathcal{R}(b_c) = \frac{\int_0^{b_c} b\,{\rm d}b\,\,{\rm AB}\,\,T_{{\rm AB}}(b)}
     {\int_0^{b_c} b\,{\rm d}b\,\, 
  \left\{1 - [1 - \sigma_{\scriptstyle{{\rm NN}}} T_{{\rm AB}}(b)]^{{\rm AB}} \right\}}.
\end{equation}

For a 5\% (10\%) centrality cut in \mbox{Pb--Pb} collisions, the yield 
$N_{\rm AB}^{\rm hard}$ is obtained by multiplying the elementary cross 
sections by $26.6~(23.7)~{\rm mb}^{-1}$.

\subsubsection*{Cross sections and yields in \PbPb~collisions at 
$\sqrtsNN=5.5~\tev$}

We used the EKS98 parameterization~\cite{EKS} of
nuclear shadowing, which corresponds to a
modification of the parton distribution functions of the nucleon
in the nucleus, $f_i^{\rm A}(x,Q^2)$, with respect to the ones of
the free nucleon, $f_i^{\rm N}(x,Q^2)$:
\begin{equation}
  \label{eq:shad}
  R_i^{\rm A}(x,Q^2) = \frac{f_i^{\rm A}(x,Q^2)}{f_i^{\rm N}(x,Q^2)}
\end{equation}
where the parton distribution functions are given as a function of the
momentum fraction $x$ carried by the parton inside the nucleon and of the QCD
scale $Q^2$ and $i = q_{\rm v},\,q_{\rm sea},\,g$ for valence quarks, 
sea quarks and gluons.
\begin{figure}
  \begin{center}
    \includegraphics[width=0.5\textwidth]{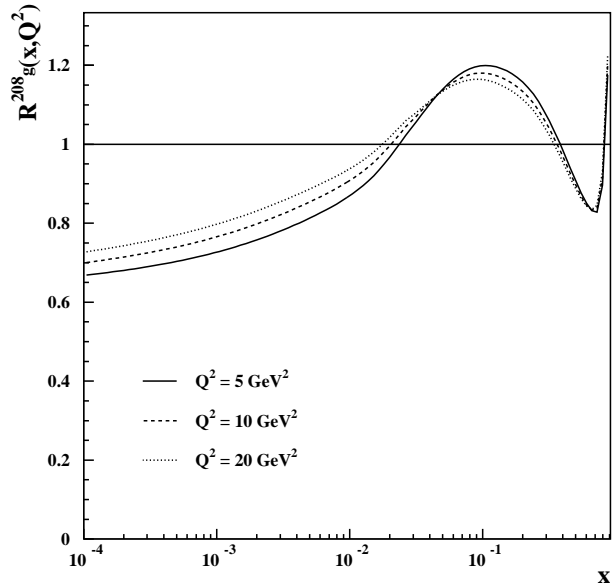}
  \end{center}
  \caption{Shadowing factor $R$ for gluons in a $^{208}$Pb nucleus as a
    function of $x$ for three values of $Q^2$.}
  \label{fig:shadowing}
\end{figure}
The shadowing factor $R_g^{\rm Pb}$ for gluons in a $^{208}$Pb nucleus is 
shown in
Fig.~\ref{fig:shadowing}. The centrality dependence of the shadowing is
weak for collisions in the considered centrality range (up to 10\% of 
$\sigma^{\rm inel}$)~\cite{emelyanov} 
and it is neglected here.

The reduction of the cross section due to shadowing amounts
to about 35\% for $\ccbar$ pairs, while it amounts only to
about 15\% for $\bbbar$ pairs, since, as seen in Section~\ref{CHAP6_5:x1x2}, 
beauty production corresponds to larger values of $x$, 
that are less affected by the shadowing suppression. 
In Section~\ref{CHAP6_5:kinematic} we will 
show how nuclear shadowing modifies the heavy quark kinematical distributions.

%The values of the parton intrinsic $k_{\rm t}$ used in the simulation
%were taken from Ref.~\cite{vogtnew}. 

Table~\ref{tab:xsecpb} reports the charm and beauty total cross
sections and yields in pp (with and without shadowing) and
\mbox{Pb--Pb} collisions at $\sqrtsNN=5.5~\tev$, as calculated with the 
HVQMNR program. 
The values shown correspond to the average of the
results obtained with MRST HO and CTEQ 5M1 parton distribution functions.
For the \PbPb~case we report the two centrality options: 5\% and 10\% of the 
total inelastic cross section. 
We have $\sigma^{\scriptstyle Q\overline{Q}}_{\rm Pb-Pb}(5\%~\sigma^{\rm inel})/\sigma^{\scriptstyle Q\overline{Q}}_{\rm Pb-Pb}(10\%~\sigma^{\rm inel})\simeq 1.12$.

\begin{table}
\caption{Total cross sections and yields for charm and beauty
  production in pp and \mbox{Pb--Pb} collisions at 
  \mbox{$\sqrtsNN=5.5~\tev$}. 
  The effect of shadowing
  is shown as the ratio $C_{\rm shad}$ of the cross section calculated
  with and without the modification of the parton distribution
  functions.
  For the \mbox{Pb--Pb} case we report the two centrality options:
  5\% and 10\% of the total inelastic cross section.}
\label{tab:xsecpb}
\begin{center}
\begin{tabular}{|cc|c|c|}
\hline
& & Charm & Beauty \\
\hline
\hline
& {\rm w/o~shadowing} & 6.64 & 0.21 \\
\cline{2-4}
\raisebox{1.5ex}[0cm][0cm]{$\sigma^{\scriptstyle Q\overline{Q}}_{{\rm pp}}\, [{\rm mb}]$} & {\rm w/~shadowing} & 4.32 & 0.18 \\
\hline
$C_{\rm shad}$ & & 0.65 & 0.84 \\
\hline
&  $5\%~\sigma^{\rm inel}$ & 45.0 & 1.79 \\
\cline{2-4}
\raisebox{1.5ex}[0cm][0cm]{$\sigma^{\scriptstyle Q\overline{Q}}_{\rm Pb-Pb} [{\rm b}]$} & $10\%~\sigma^{\rm inel}$ & 40.1 & 1.59 \\
\hline
&  $5\%~\sigma^{\rm inel}$ & 115 & 4.56 \\
\cline{2-4}
\raisebox{1.5ex}[0cm][0cm]{$N^{\scriptstyle Q\overline{Q}}_{\rm Pb-Pb}$} & $10\%~\sigma^{\rm inel}$ & 102 & 4.06 \\
\hline
\end{tabular}
\end{center}
\end{table}

\subsection{Proton--nucleus collisions}
\label{CHAP6_5:extrapolation2pPb}

For the extrapolation to \pA~collisions we use the geometrical 
Glauber-based method already described for the case of \AA~collisions. 
If we consider minimum-bias 
collisions (with no centrality selection), and we use $\rm B=1$ and 
$T_{\rm B}(\vec{s})=\delta(\vec{s})$ for the 
proton\footnote{The proton is assumed to be point-like.}, 
the total cross section for hard processes (\ref{eq:sigmaABhard}) 
becomes:
\begin{equation}
  \sigma_{{\rm pA}}^\mathrm{hard} = \sigma_{{\rm pp}}^\mathrm{hard}\cdot 2\pi \,\int_0^{\infty} b\,{\rm d}b\,\,{\rm A}\,\,T_{{\rm A}}(b) = {\rm A}\,\sigma_{{\rm pp}}^\mathrm{hard}.
\end{equation}

The number of hard processes per minimum-bias pA collision is:
\begin{equation}
  N^{\rm hard}_{\rm pA} = {\rm A}\,\sigma^{\rm hard}_{\rm pp}\bigg/\sigma^{\rm inel}_{\rm pA}.
\end{equation}

\subsubsection*{Cross sections and yields in \mbox{p--Pb} collisions 
  at $\sqrtsNN=8.8~\tev$}

Using $\rm A=208$ and $\sigma_{\rm p-Pb}^{\rm inel}=1.9$~barn~\cite{pprCh2},
the yield of $Q\overline{Q}$ pairs per minimum-bias collision is:
\begin{equation}
  N^{\scriptstyle Q\overline{Q}}_{\rm p-Pb} = \sigma^{\scriptstyle Q\overline{Q}}_{\rm pp}\cdot 0.109~{\rm mb^{-1}}.
\end{equation}

As for the \PbPb~case, the effect of nuclear shadowing was accounted 
for by using the EKS98 parameterization~\cite{EKS}. Clearly, the effect is
lower for \mbox{p--Pb}, since one of the colliding nuclei is a proton: 
the reduction of the cross sections due to nuclear shadowing is 
20\% for charm and 10\% for beauty.

The cross sections and yields for charm and beauty production in pp 
(with and without shadowing) and minimum-bias \mbox{p--Pb} collisions at 
$\sqrtsNN=8.8~\tev$ are reported in Table~\ref{tab:xsecpPb}.
The values shown correspond to the average of the
results obtained with MRST HO and CTEQ 5M1 parton distribution functions.

A summary of the production yields and of the average magnitude of nuclear
shadowing in the three considered colliding systems is presented in 
Table~\ref{tab:summarytable}.

\begin{table}
\caption{Total cross sections and yields for charm and beauty
  production in pp and \mbox{p--Pb} collisions at 
  \mbox{$\sqrtsNN=8.8~\tev$}. 
  The effect of shadowing
  is shown as the ratio $C_{\rm shad}$ of the cross section calculated
  with and without the modification of the parton distribution
  functions.} 
\label{tab:xsecpPb}
\begin{center}
\begin{tabular}{|cc|c|c|}
\hline
& & Charm & Beauty \\
\hline
\hline
& {\rm w/o~shadowing} & 9.00 & 0.30 \\
\cline{2-4}
\raisebox{1.5ex}[0cm][0cm]{$\sigma^{\scriptstyle Q\overline{Q}}_{{\rm pp}}\, [{\rm mb}]$} & {\rm w/~shadowing} & 7.16 & 0.27 \\
\hline
$C_{\rm shad}$ & & 0.80 & 0.90 \\
\hline
$\sigma^{\scriptstyle Q\overline{Q}}_{\rm p-Pb} [{\rm b}]$ & & 1.49 & 0.056 \\
\hline
$N^{\scriptstyle Q\overline{Q}}_{\rm p-Pb}$ & & 0.78 & 0.029 \\
\hline
\end{tabular}
\end{center}
\end{table}

\begin{table}
\caption{Summary table of the production yields and of the average 
         magnitude of nuclear shadowing in pp, \mbox{p--Pb} and \mbox{Pb--Pb}.}
\label{tab:summarytable}
\begin{center}
\begin{tabular}{|c|ccc|ccc|}
\hline
 & & Charm & & & Beauty & \\
\hline
system & pp & p--Pb & Pb--Pb & pp & p--Pb & Pb--Pb \\
centrality & min.-bias & min.-bias & centr. (5\%) & min.-bias & min.-bias & centr. (5\%) \\ 
$\sqrtsNN$ & $14~\tev$ & $8.8~\tev$ & $5.5~\tev$ &  $14~\tev$ & $8.8~\tev$ & $5.5~\tev$ \\
\hline
\hline
$N^{\scriptstyle Q\overline{Q}}$/ev & 0.16 & 0.78 & 115 & 0.0072 & 0.029 & 4.56 \\
$C_{\rm shad}$ & 1 & 0.80 & 0.65 & 1 & 0.90 & 0.84 \\
\hline
\end{tabular}
\end{center}
\end{table}

\section{Heavy quark kinematical distributions}
\label{CHAP6_5:kinematic}

\begin{figure}
  \begin{center}
    \includegraphics[width=\textwidth]{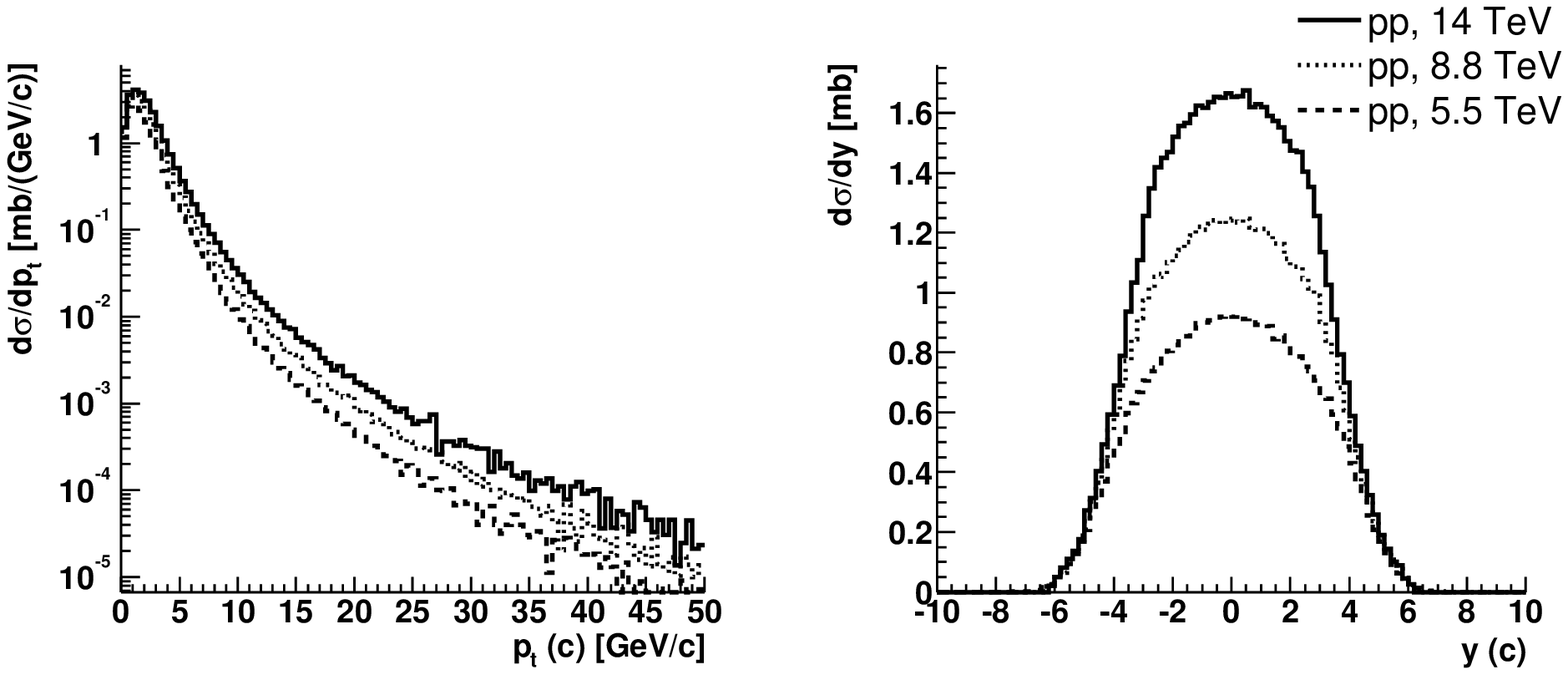}
    \includegraphics[width=\textwidth]{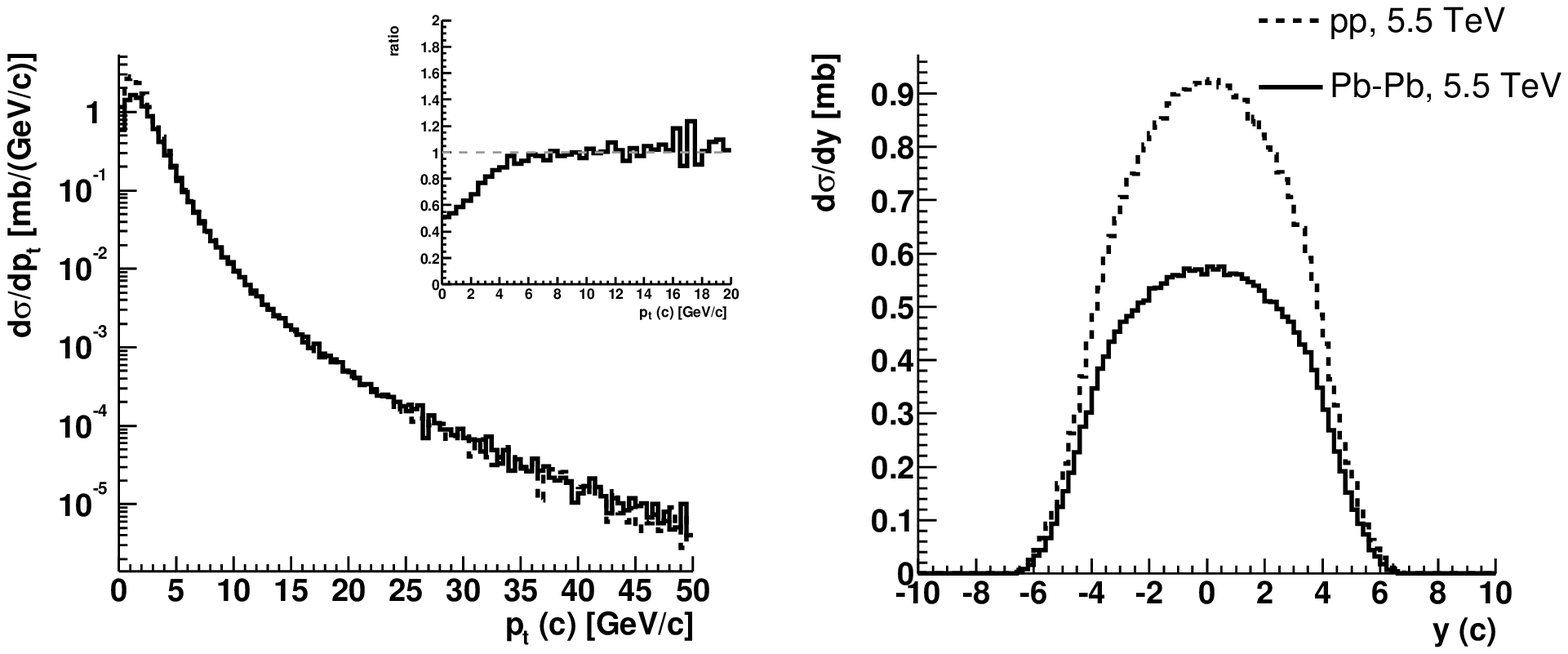}
    \includegraphics[width=\textwidth]{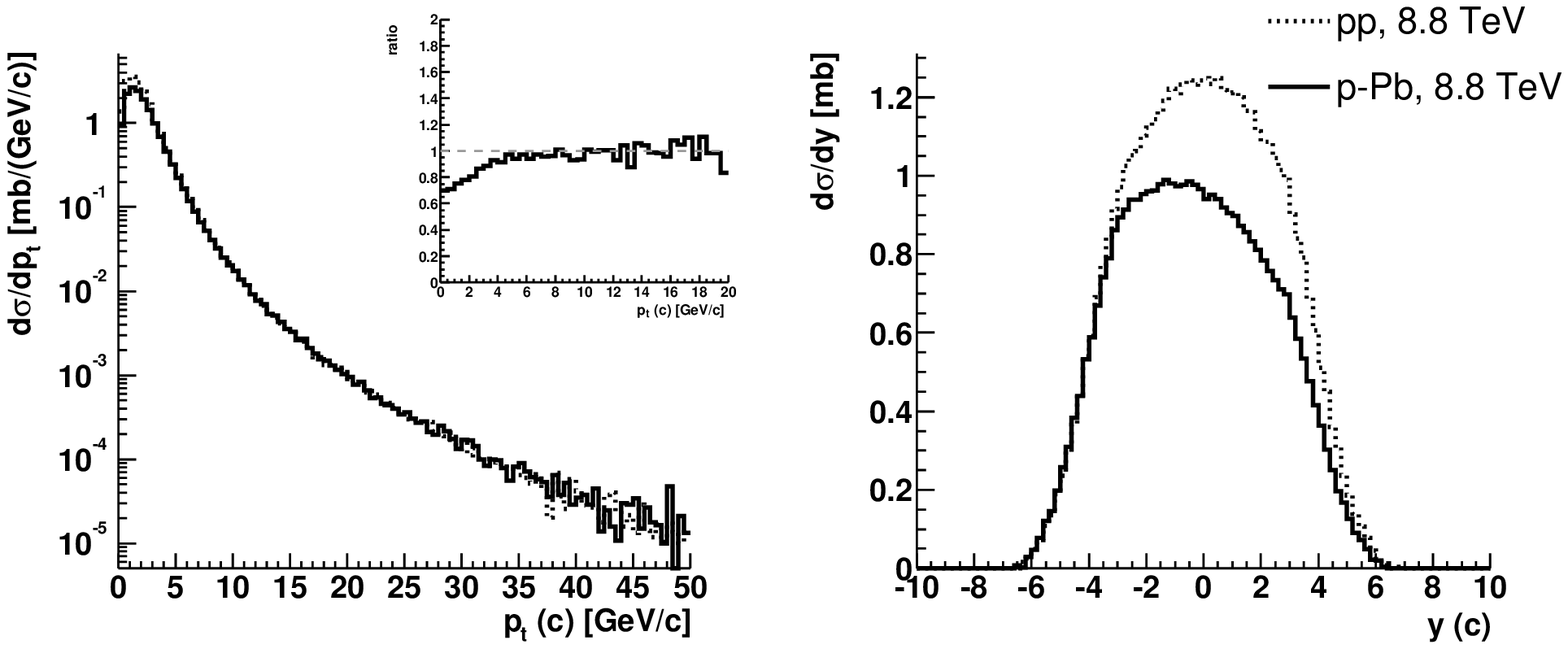}
  \end{center}
  \caption{Inclusive c quark $\pt$ and rapidity distributions obtained 
           from the HVQMNR program. 
           The distributions for \PbPb~and \pPb~are 
           normalized to the cross sections per \NN~collision and they 
           include the effects of nuclear shadowing and intrinsic 
           $k_{\rm t}$ broadening.}
  \label{fig:ckine}
\end{figure}

\begin{figure}
  \begin{center}
    \includegraphics[width=\textwidth]{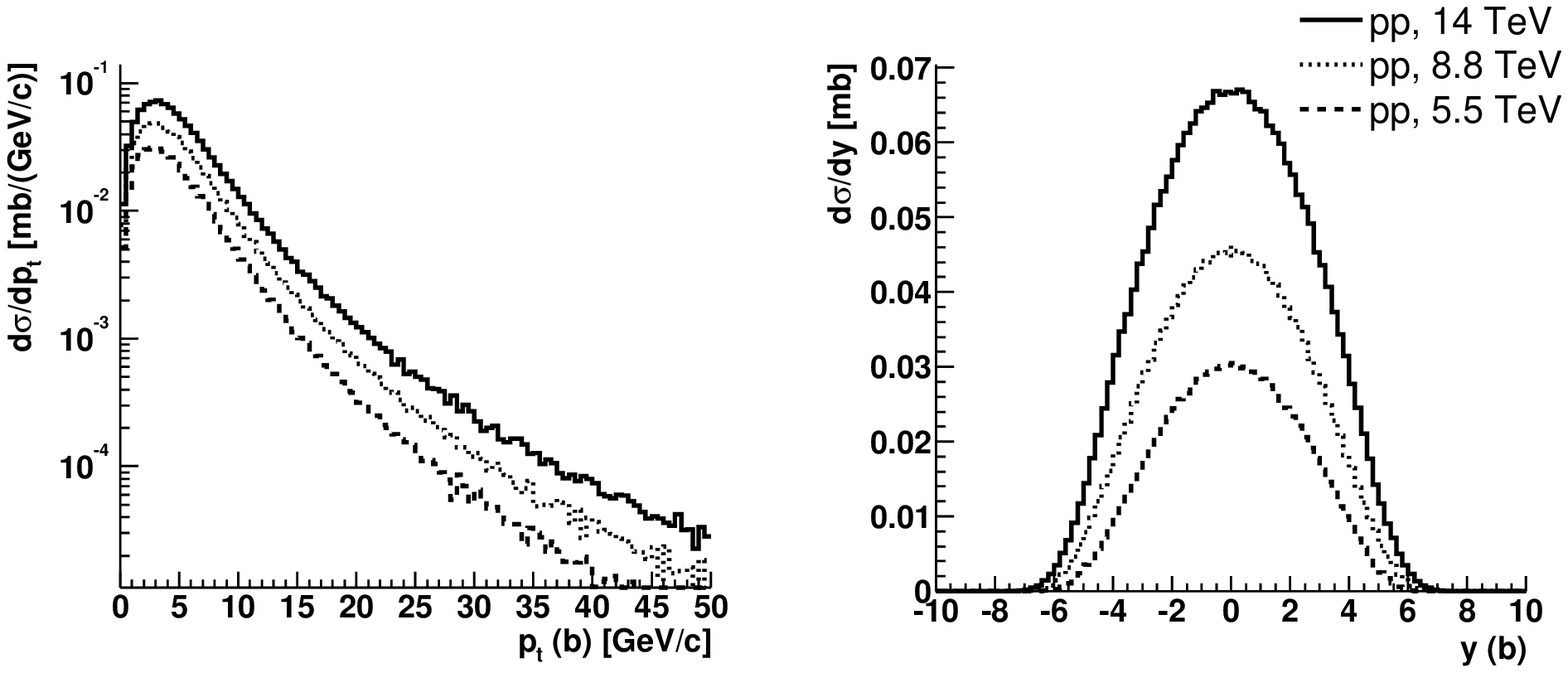}
    \includegraphics[width=\textwidth]{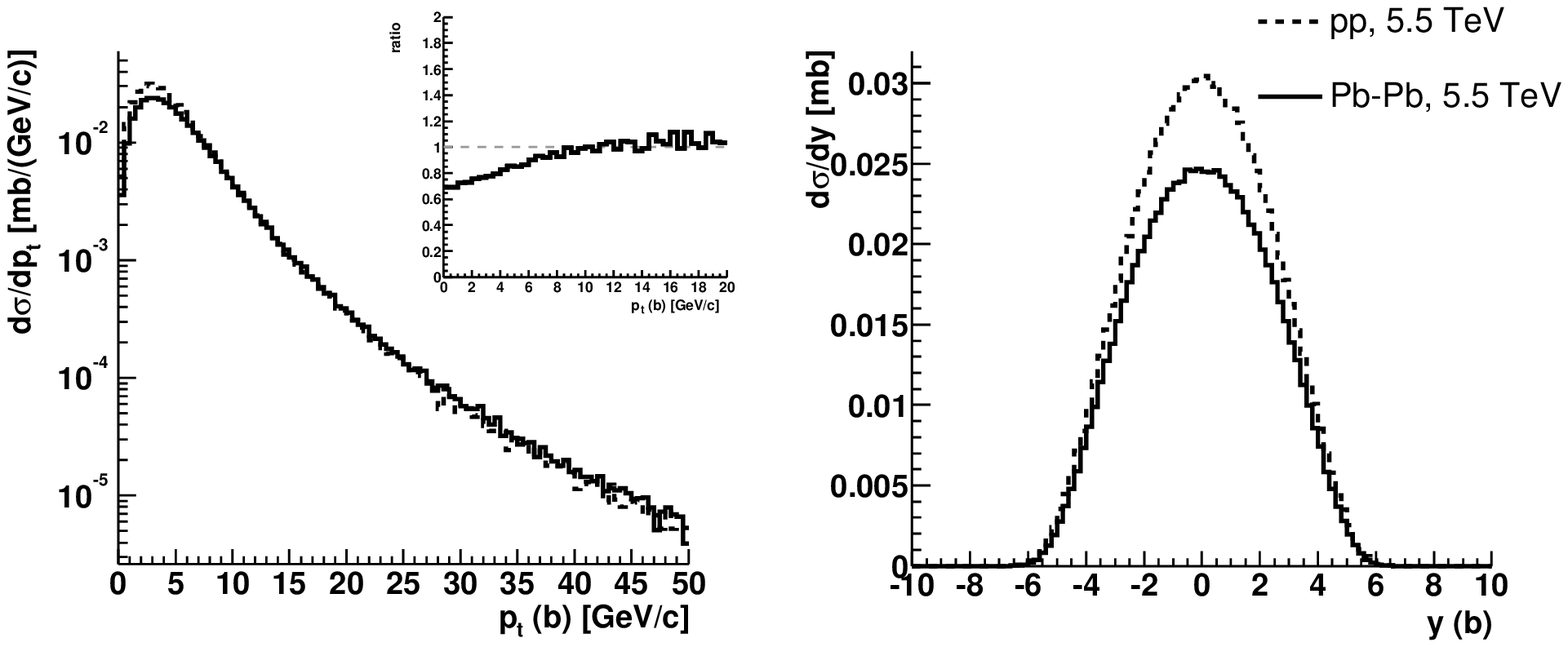}
    \includegraphics[width=\textwidth]{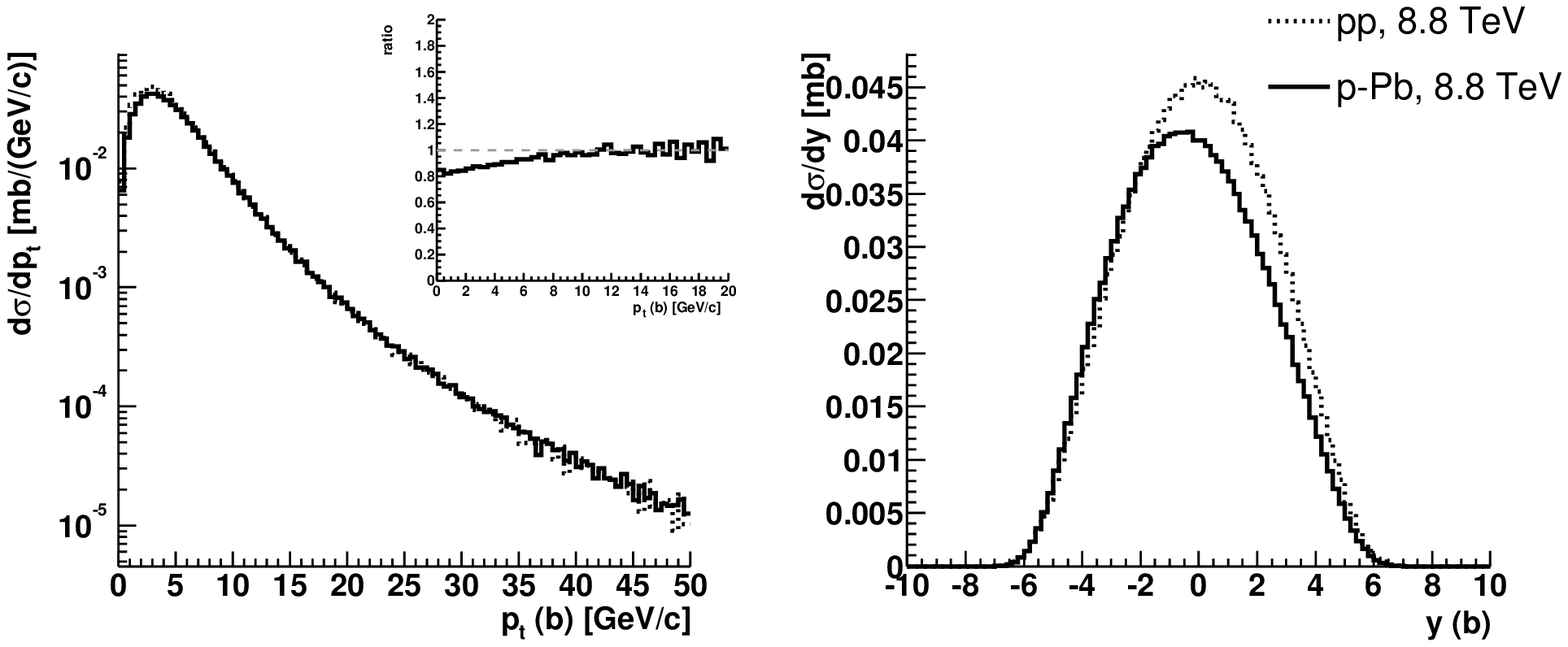}
  \end{center}
  \caption{Inclusive b quark $\pt$ and rapidity distributions obtained 
           from the HVQMNR program. 
           The distributions for \PbPb~and \pPb~are 
           normalized to the cross sections per \NN~collision and they 
           include the effects of nuclear shadowing and intrinsic 
           $k_{\rm t}$ broadening.}
  \label{fig:bkine}
\end{figure}

Figures~\ref{fig:ckine} and~\ref{fig:bkine} present the transverse momentum 
and rapidity distributions, obtained using the NLO pQCD program 
HVQMNR, for c and b quarks, respectively.
The distributions for \PbPb~and \pPb~are normalized to the cross 
section per \NN~collision. 

We used the CTEQ 4M~\cite{cteq4} set of PDF.
We verified that the results given by this set 
lie in between the ones obtained with the more recent CTEQ 5 and MRST sets 
for all the relevant kinematical quantities~\cite{yelrepHardProbes}. 
For the other parameters
to values specified in Section~\ref{CHAP6_5:XsecNN} were used: 
$m_{\rm c}=1.2~\gev$, 
$\mu_R=\mu_F=2\,\mu_0$ for charm and $m_{\rm b}=4.75~\gev$, 
$\mu_R=\mu_F=\mu_0$ for beauty.
Nuclear shadowing is included via the EKS98 
parameterization~\cite{EKS}. 
The parton intrinsic $k_{\rm t}$ is sampled from a 
Gaussian distribution with mean 0 and $\sigma$ ($=\sqrt{\av{k_{\rm t}^2}}$) 
equal to 1, 1.16, $1.30~\gev/c$ 
for charm production in pp, p--Pb and Pb--Pb, respectively,
and equal to 1, 1.60, $2.04~\gev/c$ 
for beauty production in pp, p--Pb and Pb--Pb, respectively. These values 
are taken from Ref.~\cite{vogtnew}. The same parameters are used also in the 
calculations shown in the next section.

In the case of \pPb~events the rapidity distribution in the centre-of-mass 
frame is plotted; the rapidity distribution in the laboratory frame 
would be shifted by $\Delta y=0.47$.

We notice that the $\pt$ distributions for pp collisions at 5.5, 8.8 and 14~TeV
 (top-left panel) have essentially the same shape.

The comparison of the $\pt$ distributions for pp and \PbPb~(and for 
pp and \pPb) at the same centre-of-mass energy shows that
nuclear shadowing affects heavy quark production only for relatively 
low transverse momenta ($\pt<5$-$6~\gev/c$ with EKS98), 
where the $\QQbar$ pairs
are produced by low-$x$ gluons. This is clearly seen in the ratios of the 
distributions, reported in the insets. The value for the upper limit, 
$\approx 5~\gev/c$,
of the $\pt$-region affected by the shadowing in \PbPb~collisions
can be cross-checked with the following simple estimate:
for the back-to-back production of a $\ccbar$ pair at central 
rapidity, with transverse momenta 
$\pt^{\rm c}=\pt^{\rm \overline{c}}=5~\gev/c$, we have $Q\simeq 2\,\pt=10~\gev$
and $x\simeq Q/\sqrtsNN=10/5500\simeq 2\cdot 10^{-3}$; for these values of 
$x$ and $Q$, the EKS98 parameterization gives $R^{\rm Pb}_g$ 
(defined in Eq.~(\ref{eq:shad})) $\simeq 90\%$.
This suppression is already quite small and it is partially compensated 
by the $k_{\rm t}$ broadening.

A relevant feature of $\QQbar$ production in \pPb~collisions
is a depletion in the forward region (where the proton goes) of the rapidity 
distributions. This effect is due to the shadowing which is biased toward
forward rapidities, where the smallest $x$ values in the Pb nucleus 
are probed. 

\section{Heavy quark production in Monte Carlo event generators}
\label{CHAP6_5:generators}

\begin{figure}[!t]
  \begin{center}
    \includegraphics[width=.65\textwidth]{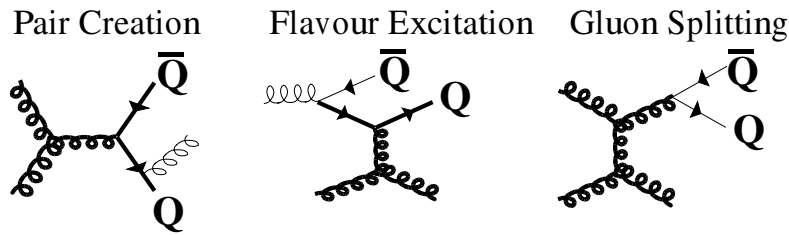}
  \caption{Some of the PYTHIA processes defined as pair creation, flavour
    excitation and gluon splitting. The thick lines correspond to the
    hard process, the thin ones to the initial- or final-state parton shower.}
  \label{fig:processes}
  \end{center}
\end{figure}

The program used for the NLO calculations reported in
the previous sections is
not well suited to be included in a simulation, since it is not
an event generator and it does not provide parton kinematics, but only 
inclusive distribution.
On the other hand, widely used event
generators, like PYTHIA~\cite{pythia} and HERWIG~\cite{herwig}, 
are exact only at leading order, when only the {\sl pair production} processes,
$q\overline{q}\to Q\overline{Q}$ and $gg\to Q\overline{Q}$ 
(see Fig.~\ref{fig:processes}), are included. 
Higher-order contributions are included in these generators
in the parton shower approach (see e.g. Ref.~\cite{norrbin}). 
This model is not exact at next-to-leading order, but it reproduces 
some aspects of the multiple-parton-emission phenomenon.
In the following we will concentrate on the PYTHIA event generator; 
the version we have used is PYTHIA 6.150. We have also investigated 
heavy quark production in HERWIG, observing an incorrect behaviour
in the final kinematical distributions of both c and b quarks. We, therefore, 
concluded that HERWIG is not suitable for heavy quark simulations at LHC
energies. More details can be found in Ref.~\cite{yelrepHardProbes}.  

%In PYTHIA, the processes are classified
%according to the number of heavy quarks in the final state of the hard
%process, where the hard process is defined as the one with the highest
%virtuality $Q^2$ (i.e. highest invariant mass of the outgoing parton pair, 
%as defined in equation (\ref{eq:sx1x2M2})). There are three classes
%of processes:
%\begin{description}
%\item[pair creation:] the hard process is one of the leading-order
%  graphs ($gg \rightarrow Q\overline{Q}$, $q\overline{q}
%  \rightarrow Q\overline{Q}$); its final state contains two heavy
%  quarks;
%\item[flavour excitation:] an incoming heavy quark is put on mass shell
%  by scattering on a parton of the other beam: $qQ\rightarrow qQ$ or
%  $gQ\rightarrow gQ$; the incoming heavy quark must come from a $g
%  \rightarrow Q\overline{Q}$ in the initial state shower; this
%  process is characterized by one heavy quark in the final state of
%  the hard scattering;
%\item[gluon splitting:] no heavy flavour is involved in the hard
%  scattering, but a $Q\overline{Q}$ pair is produced in the initial or
%  final state showers from a $g \rightarrow Q\overline{Q}$
%  branching.
%\end{description}
%Figure~\ref{fig:processes} shows some topologies belonging to the
%processes specified above.

\begin{figure}[!b]
  \begin{center}
    \includegraphics[width=\textwidth]{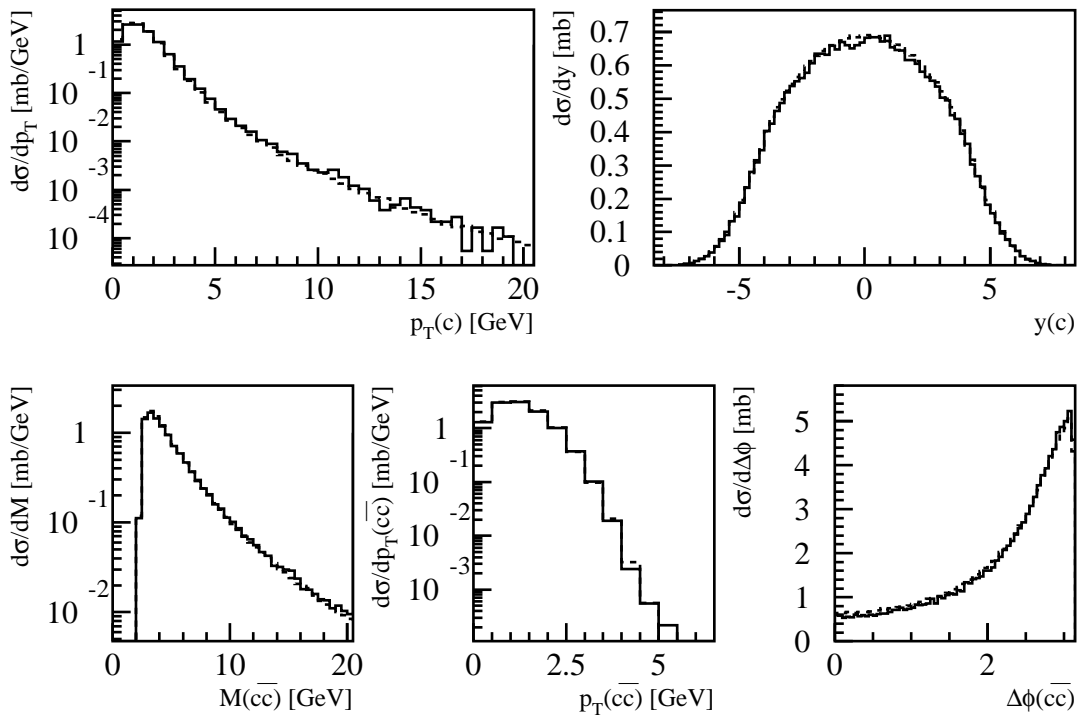}
  \end{center}
  \caption{Comparison between PYTHIA results (solid histogram) for the LO
    process $gg\to c\overline{c}$, without parton shower, 
    and corresponding HVQMNR prediction (dashed histogram). 
    The centre of mass energy is $\sqrt{s}=5.5~\tev$.}
  \label{fig:cmpCharmLO}
\end{figure}

In PYTHIA, the processes giving rise to contributions above
leading order, 
like (see Fig.~\ref{fig:processes}) 
{\sl flavour excitation}, $qQ\rightarrow qQ$ and
$gQ\rightarrow gQ$, and {\sl gluon splitting}, $g\rightarrow
Q\overline{Q}$, are calculated using a massless matrix element. As a
consequence the cross sections for these processes diverge as
$\pt^{\rm hard}$ vanishes\footnote{$\pt^{\rm hard}$ is defined as the
transverse momentum of the outgoing quarks in the rest frame of the
hard interaction.}. These divergences are regularized by putting a lower
cut-off on $\pt^{\rm hard}$. The value of the minimum
$\pt^{\rm hard}$ cut has a large influence on the heavy flavour cross
section at low $\pt$, a region of prime interest for ALICE physics and
covered by the ALICE acceptance. Our approach was, therefore, to
tune the PYTHIA parameters in order to reproduce as well as possible the NLO
predictions (HVQMNR). 
We used PYTHIA with the option MSEL=1, that allows to switch 
on one by one the different processes (see the Appendix 
for more details). 
The main parameter we tuned is the lower $\pt^{\rm hard}$
limit. In this procedure we compared the following distributions of the
bare quarks:
\begin{itemize}
\item inclusive $\pt$ and rapidity distributions of the quark (antiquark);
\item mass of the pair: $M_{Q\overline{Q}} =
  \sqrt{(E_Q+E_{\overline{Q}})^2-(\vec{p}_Q+\vec{p}_{\overline{Q}})^2}$,
  where $E_Q = \sqrt{m_Q^2 + p_Q^2}$ is the quark energy;
\item $\pt$ of the pair, defined as the projection on the plane normal
  to the beam axis of the $Q\overline{Q}$ total momentum;
\item angle $\Delta\phi$ between the quark and the antiquark in the
  plane normal to the beam axis.
\end{itemize}

In the simulations for \mbox{Pb--Pb} collisions at $\sqrtsNN=5.5~\tev$ 
the parton distribution functions used are the
CTEQ 4, modified for nuclear shadowing using the EKS98~\cite{EKS}
parameterization. 

\begin{figure}
  \begin{center}
    \includegraphics[width=0.9\textwidth]{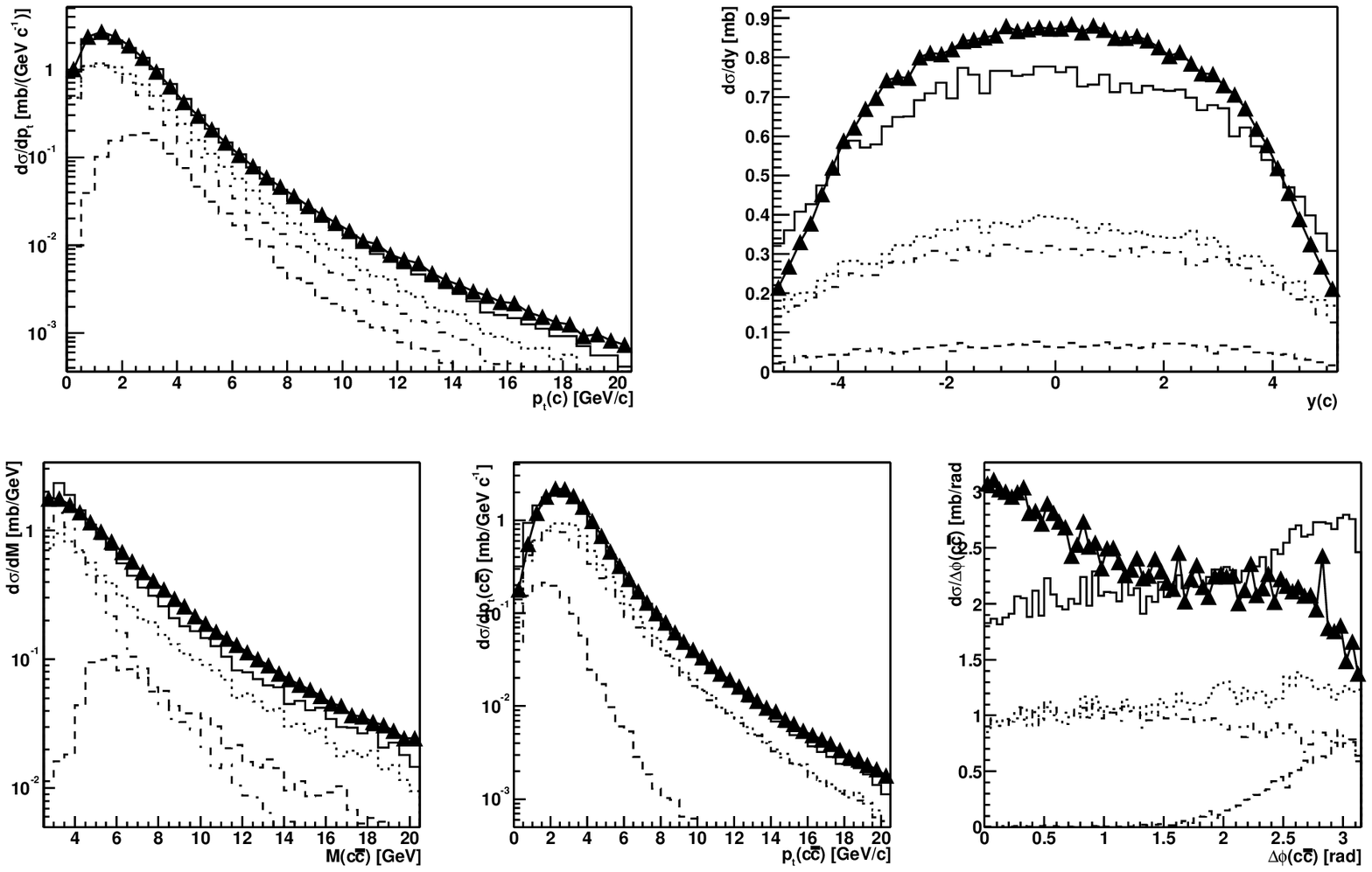}
%  \end{center}
  \caption{Comparison between charm production in \mbox{Pb--Pb} collisions at 
    $\sqrtsNN=5.5~\tev$ in the HVQMNR NLO calculation
    and in PYTHIA with parameters tuned as
    described in the text. The triangles show the NLO calculation, the
    solid histogram corresponds to the PYTHIA total production. The
    individual PYTHIA contributions are pair production (dashed),
    flavour excitation (dotted) and gluon splitting (dot-dashed).}
  \label{fig:charmPbPbPyMNR}
%\end{figure}
%\begin{figure}
%  \begin{center}
    \includegraphics[width=0.9\textwidth]{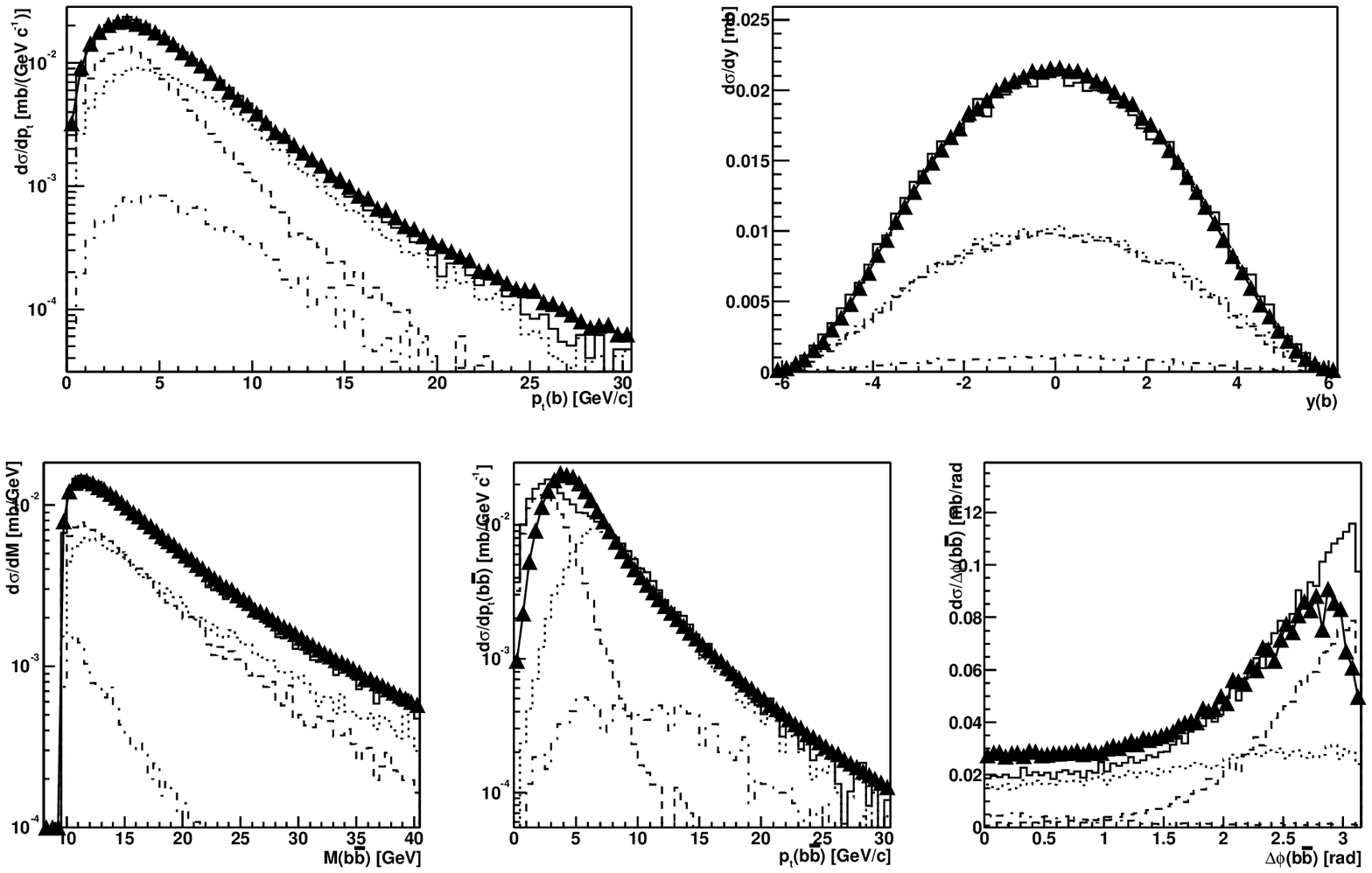}
  \caption{Equivalent of Fig.~\ref{fig:charmPbPbPyMNR} for beauty production.}
  \label{fig:beautyPbPbPyMNR}
  \end{center}
\end{figure}

\begin{figure}
  \begin{center}
    \includegraphics[width=0.9\textwidth]{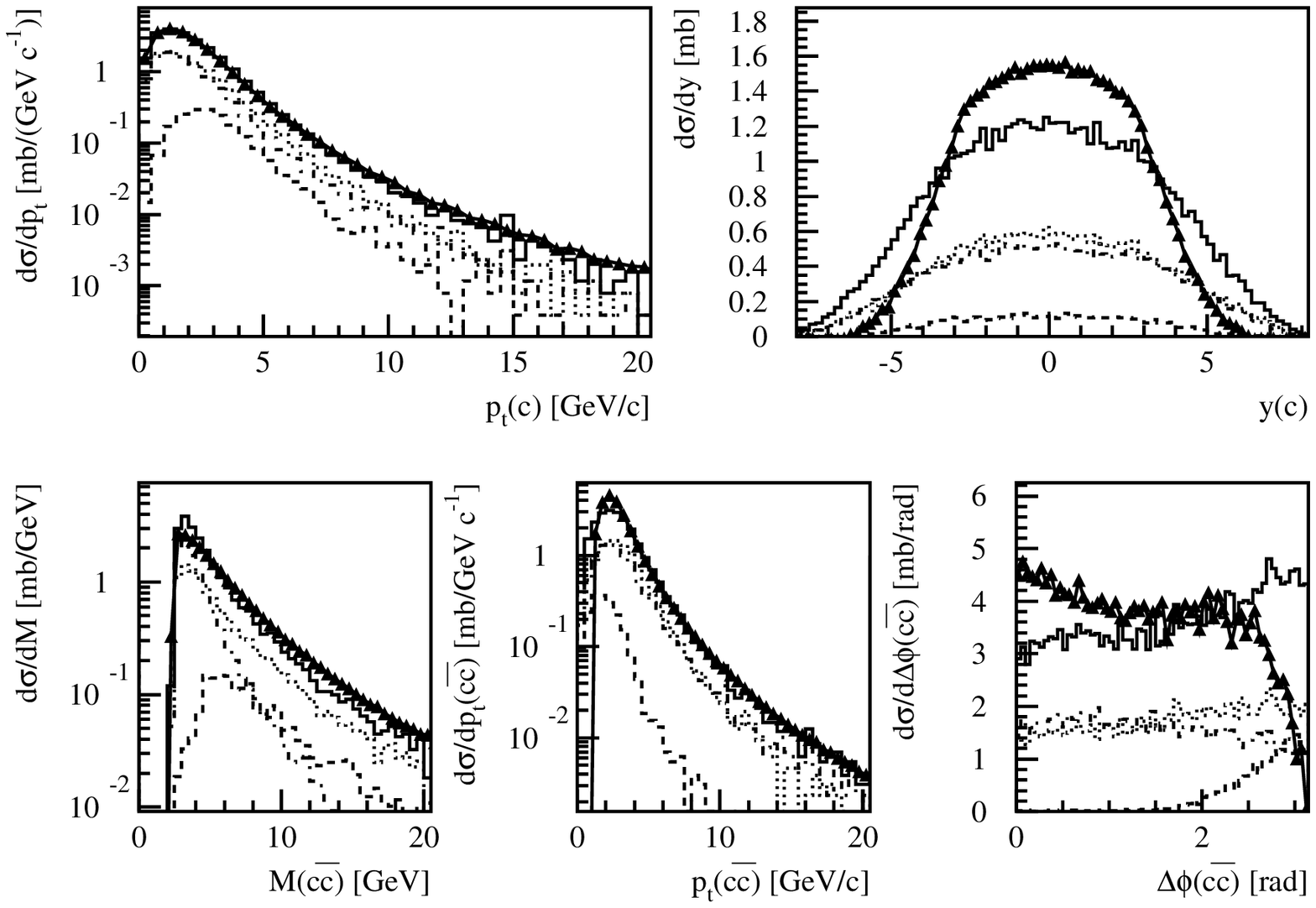}
%  \end{center}
  \caption{Comparison between charm production in pp collisions at 
    $\sqrt{s}=14~\tev$ in the HVQMNR NLO calculation
    and in PYTHIA with parameters tuned as
    described in the text. The triangles show the NLO calculation, the
    solid histogram corresponds to the PYTHIA total production. The
    individual PYTHIA contributions are pair production (dashed),
    flavour excitation (dotted) and gluon splitting (dot-dashed).}
  \label{fig:charmPpPyMNR}
%\end{figure}
%\begin{figure}
%  \begin{center}
    \includegraphics[width=0.9\textwidth]{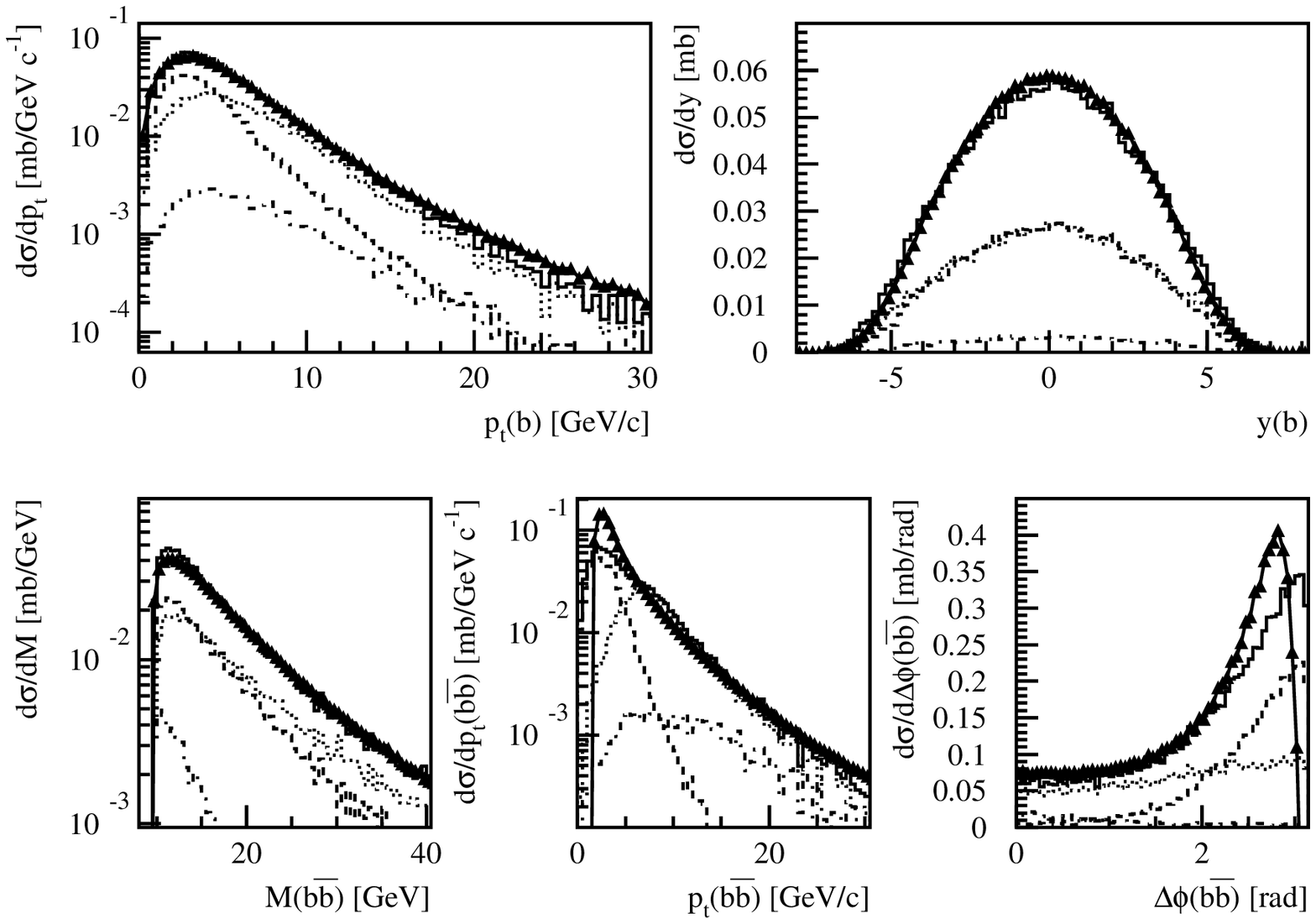}
  \caption{Equivalent of Fig.~\ref{fig:charmPpPyMNR} for beauty production.}
  \label{fig:beautyPpPyMNR}
  \end{center}
\end{figure}

Before presenting the results of the tuning of PYTHIA to reproduce the pQCD 
results at NLO, we show that, with the same input parameters, PYTHIA and 
the pQCD calculation in HVQMNR give exactly the same kinematical 
distributions for the LO process $gg\to Q\overline{Q}$.
The comparison is reported in Fig.~\ref{fig:cmpCharmLO} 
for $\ccbar$ production in pp collisions at $\sqrt{s}=5.5~\tev$;
the normalization is set to the value of the cross section obtained for 
\pp~without shadowing (first row of Table~\ref{tab:xsecpb}) and the 
PYTHIA results are scaled to this number.
 
The results of the tuning to pQCD at NLO 
are shown in Figs.~\ref{fig:charmPbPbPyMNR}
and~\ref{fig:beautyPbPbPyMNR}, where the distributions from PYTHIA and the
NLO calculation are compared.  In this case the overall normalization is 
set to the value of the cross sections obtained for \pp~with shadowing  
(second row of Table~\ref{tab:xsecpb}). Despite the
fundamental differences between the two models, the agreement is
relatively good. However, significant discrepancies are present,
especially in the $\Delta\phi$ distribution for \ccbar~pairs.

A similar tuning of the PYTHIA event generator was done also for the 
production of $\ccbar$ and $\bbbar$ pairs in pp collisions at 
$\sqrt{s}=14~\tev$.
The same set of parton distribution functions (CTEQ 4) was used, 
without the modification for nuclear shadowing. 
Results are shown in 
Figs.~\ref{fig:charmPpPyMNR} and~\ref{fig:beautyPpPyMNR}. The largest 
difference with the results obtained for the \mbox{Pb--Pb} case is a worse 
description of the rapidity distribution of charm quarks. This is due 
to a feature of the parameterizations of the parton distribution functions:
most of them, including CTEQ 4, are valid only
down to $x=10^{-5}$; below this value the behaviour depends on 
the implementation of the specific parameterization but has no physical 
meaning (e.g. for the CTEQ 4 the gluon density $g(x)$ is kept constant at 
$g(10^{-5})$). The rapidity range in which the evolution of the parton 
distribution functions is reliable depends on the c.m.s. energy; for 
charm production at $\sqrt{s}=5.5~\tev$ (14~$\tev$) this range is 
found to be $|y|<4.3$ ($|y|<3.4$), using equation~(\ref{eq:yx1x2}) with 
$x_1>10^{-5}$ and $x_2>10^{-5}$. This feature is not present in the latest
CTEQ set of PDF, CTEQ 6~\cite{cteq6}, which is evolved down to $x=10^{-6}$.

The values of the PYTHIA parameters obtained from the tuning are reported 
in the Appendix.

\section{Hadron yields and distributions}
\label{CHAP6_5:hadr}

For the hadronization of heavy quarks we use the default Lund string
fragmentation model~\cite{norrbin} included in PYTHIA (JETSET package).
The total yield and the rapidity density d$N$/d$y$ in the central region 
for hadrons with open 
charm and beauty in \mbox{Pb--Pb} at 5.5~$\tev$~($5\%~\sigma^{\rm tot}$ 
centrality selection), pp at 14~$\tev$ and \mbox{p--Pb} at 
$8.8~\tev$ are summarized 
in Tables~\ref{tab:hadyieldsPbPb},~\ref{tab:hadyieldspp} 
and~\ref{tab:hadyieldspPb}, respectively.
The rapidity densities are calculated in $-1<y_{\rm lab}<1$, corresponding to 
$-1.47<y_{\rm c.m.s.}<0.53$ for \mbox{p--Pb} 
and $-0.53<y_{\rm c.m.s.}<1.47$ for \mbox{Pb--p}. 
No dependence of the relative hadron abundances on the centre-of-mass 
energy is observed. 

\begin{table}
  \caption{Total yield and average rapidity density for $|y|<1$ for
    hadrons with charm and beauty in \mbox{Pb--Pb} collisions at 
    $\sqrtsNN=5.5~\tev$. The values reported correspond to a centrality
    selection of $5\%~\sigma^{\rm tot}$.}
  \label{tab:hadyieldsPbPb}
  \begin{center}
  \begin{tabular}{|ccc|ccc|}
\hline
  Particle & Yield & $\langle$d$N$/d$y\rangle_{|y|<1}$ & Particle & Yield &
    $\langle$d$N$/d$y\rangle_{|y|<1}$ \\
\hline
\hline
${\rm D^0}$&   68.9&   6.87&${\rm B^0}$&   1.86&  0.273\\
${\rm \overline{D}^0}$&   71.9&   6.83&${\rm \overline{B}^0}$&   1.79&  0.262\\
${\rm D^+}$&   22.4&   2.12&${\rm B^+}$&   1.82&  0.251\\
${\rm D^-}$&   22.2&   2.00&${\rm B^-}$&   1.83&  0.270\\
${\rm D_s^+}$&   14.1&   1.30&${\rm B_s^0}$&   0.53&  0.077\\
${\rm D_s^-}$&   12.7&   1.22&${\rm \overline{B}_s^0}$&   0.53&  0.082\\
${\rm \Lambda_c^+}$&    9.7&   1.18&${\rm \Lambda_b^0}$&   0.36&  0.050\\
${\rm \overline{\Lambda_c}^-}$&    8.2&   0.85&${\rm \overline{\Lambda_b}^0}$&
    0.31&  0.047\\
\hline
  \end{tabular}
  \end{center}
%\end{table}

%\begin{table}
  \caption{Total yield and average rapidity density for $|y|<1$ for
    hadrons with charm and beauty in pp collisions at 
    $\sqrt{s}=14~\tev$.}
  \label{tab:hadyieldspp}
  \begin{center}
  \begin{tabular}{|ccc|ccc|}
\hline
  Particle & Yield & $\langle$d$N$/d$y\rangle_{|y|<1}$ & Particle & Yield &
    $\langle$d$N$/d$y\rangle_{|y|<1}$ \\
\hline
\hline
${\rm D^0}$ &   0.0938 & 0.0098 & ${\rm B^0}$ & 0.00294 & 0.00043\\
${\rm \overline{D}^0}$ & 0.0970 & 0.0098 & ${\rm \overline{B}^0}$ & 0.00283 & 0.00041\\
${\rm D^+}$ & 0.0297 & 0.0029&${\rm B^+}$ & 0.00287 & 0.00040\\
${\rm D^-}$ & 0.0290 & 0.0029&${\rm B^-}$ & 0.00289 & 0.00043\\
${\rm D_s^+}$ & 0.0186 & 0.0018&${\rm B_s^0}$ & 0.00084 & 0.00012\\
${\rm D_s^-}$ & 0.0176 & 0.0020&${\rm \overline{B}_s^0}$ & 0.00084 & 0.00013\\
${\rm \Lambda_c^+}$ & 0.0113 & 0.0013 & ${\rm \Lambda_b^0}$ & 0.00057 & 0.00008\\
${\rm \overline{\Lambda_c}^-}$ &    0.0110 & 0.0013 & ${\rm \overline{\Lambda_b}^0}$ & 0.00049 & 0.00008\\
\hline
  \end{tabular}
  \end{center}
%\end{table}

%\begin{table}
  \caption{Total yield and average rapidity density for $|y_{\rm lab}|<1$ for
    hadrons with charm and beauty in \mbox{p--Pb} collisions at 
    $\sqrtsNN=8.8~\tev$.}
  \label{tab:hadyieldspPb}
  \begin{center}
  \begin{tabular}{|ccc|ccc|}
\hline
  Particle & Yield & $\langle$d$N$/d$y\rangle_{|y_{\rm lab}|<1}$ & 
  Particle & Yield & $\langle$d$N$/d$y\rangle_{|y_{\rm lab}|<1}$ \\
\hline
\hline
${\rm D^0+\overline{D}^0}$ &   0.926 & 0.096 & ${\rm B^0+\overline{B}^0}$ & 0.0221 & 0.0030\\
${\rm D^++D^-}$ & 0.293 & 0.030 & ${\rm B^++B^-}$ & 0.0221 & 0.0030\\
${\rm D_s^++D_s^-}$ & 0.176 & 0.018&${\rm B_s^0+\overline{B}_s^0}$ & 0.0064 & 0.0009\\
${\rm \Lambda_c^++\overline{\Lambda_c}^-}$ & 0.118 & 0.012 & ${\rm \Lambda_b^0+\overline{\Lambda_b}^0}$ & 0.0041 & 0.0005\\
\hline
  \end{tabular}
  \end{center}
\end{table}

It is interesting to notice the large ratio of the neutral-to-charged 
D meson yields: $N({\rm D^0})/N({\rm D^+})\simeq 3.1$. 
In PYTHIA, charm quarks are 
assumed to fragment to D (spin singlets: $J=0$) and D$^*$ (spin triplets: 
$J=1$) mesons according to the number of available spin states; therefore,
 $N({\rm D^0}):N({\rm D^+}):N({\rm D^{*0}}):N({\rm D^{*+}})=1:1:3:3$.
Then, the resonances D$^*$ are decayed to D mesons according to the 
branching ratios. The difference between neutral and charged D mesons arises 
here: due to the slightly larger ($\approx 4~\mev$) mass of the D$^+$, the 
D$^{*+}$ decays preferably to $\Dz$ and the D$^{*0}$ decays exclusively to 
$\Dz$. We have~\cite{pdg}:
\begin{equation}
\begin{array}{rcl}
\displaystyle
\frac{N({\rm D^0})}{N({\rm D^+})}&=&\displaystyle\frac{N({\rm D^0_{primary}})+N({\rm D^{*+}})\times BR({\rm D^{*+}\to D^0})+N({\rm D^{*0}})\times BR({\rm D^{*0}\to D^0})}{N({\rm D^+_{primary}})+N({\rm D^{*+}})\times BR({\rm D^{*+}\to D^+})+N({\rm D^{*0}})\times BR({\rm D^{*0}\to D^+})}\\
&=&\displaystyle\frac{1+3\times 0.68+3\times 1}{1+3\times 0.32+3\times 0}\\
&=& 3.08.
\end{array}
\end{equation}  

We chose to use the relative abundances given by PYTHIA, although, 
experimentally, the fraction $\Dz/{\rm D^+}$ was found to be lower than 3.
The value measured in $e^+e^-$ collisions at LEP by the ALEPH Collaboration 
is $\approx 2.4$~\cite{alephD}. This would reduce by about 6\% the 
expected yield for the $\Dz$ mesons.

\begin{figure}[!t]
  \begin{center}
    \includegraphics[width=0.49\textwidth]{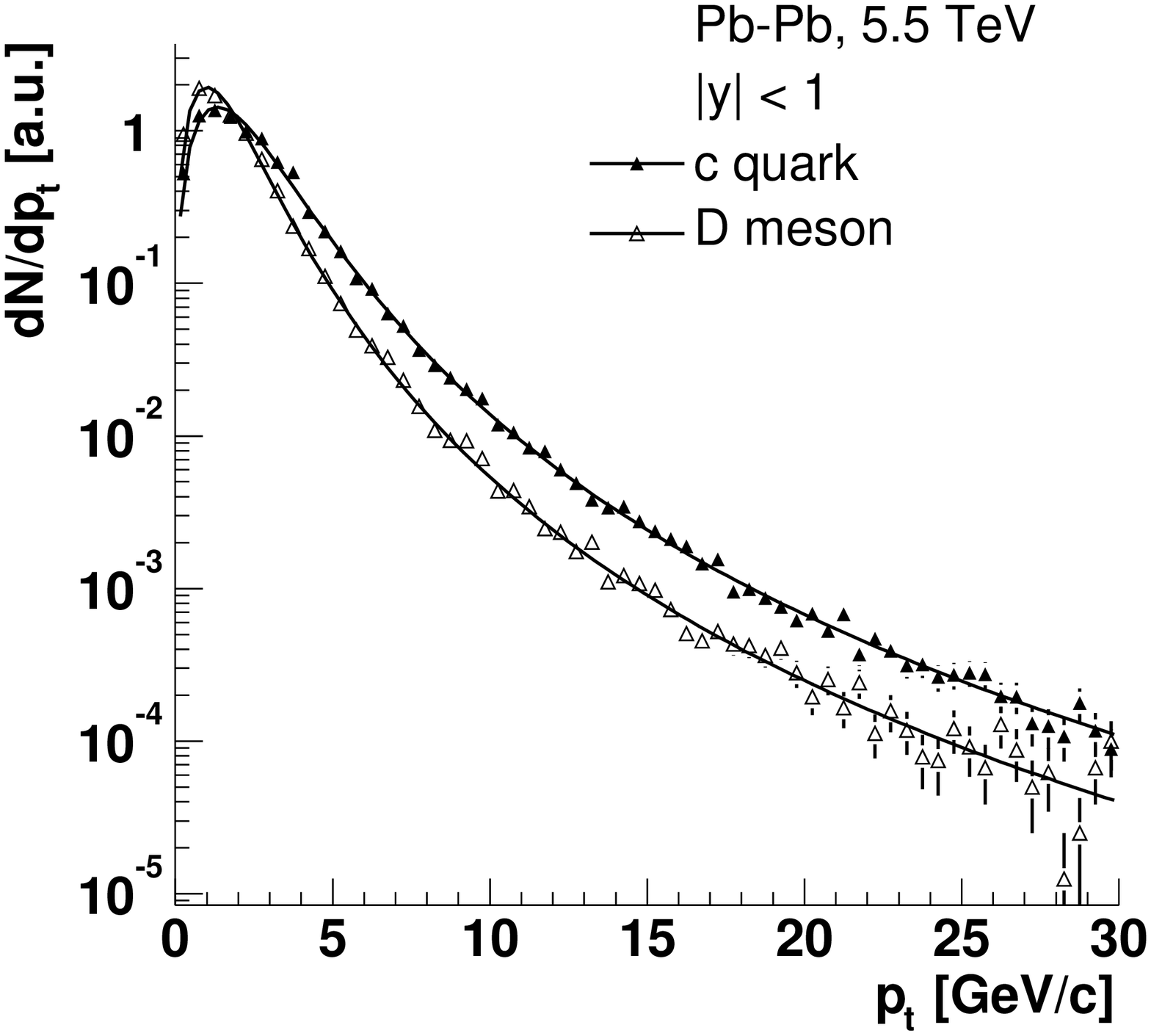}
    \includegraphics[width=0.49\textwidth]{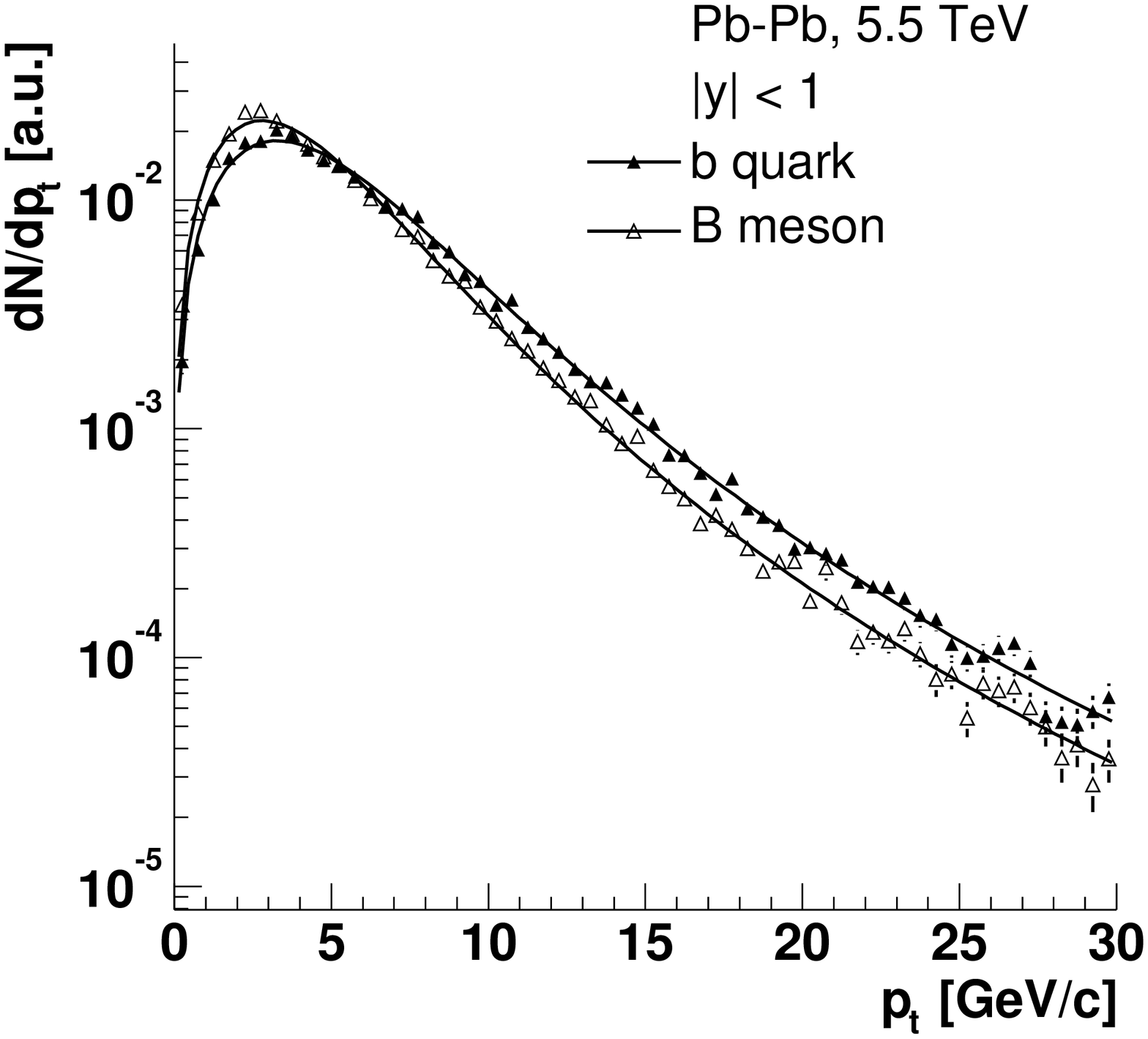}
  \end{center}
  \caption{Transverse momentum distributions at mid-rapidity for heavy quarks 
           and mesons in \mbox{Pb--Pb} at 5.5 TeV. The distributions are 
           normalized to the same integral in order to compare their
           shapes.}
  \label{fig:ptDcBb}
\end{figure}
 
Figure~\ref{fig:ptDcBb} presents the transverse
momentum distributions at mid-rapidity ($|y|<1$) 
for c quarks and D mesons (left panel) and  
for b quarks and B mesons (right panel), in \mbox{Pb--Pb} at 5.5~$\tev$. 
For $\pt>0$ and $|y|<1$, we have, on average, 
$\pt^{\rm D}\simeq 0.75\,\pt^{\rm c}$
and $\pt^{\rm B}\simeq 0.85\,\pt^{\rm b}$. 
The shape of the transverse momentum distributions for D and B mesons 
was fitted to the following expression:
\begin{equation}
  \label{eq:fitMesons}
  \frac{1}{\pt}\frac{{\rm d}N}{{\rm d}\pt}\propto\left[1+\left(\frac{\pt}{\pt^0}\right)^2\right]^{-n}.
\end{equation}
The $\pt$ distributions were studied also for pp at 
14~TeV and for \pPb~at 8.8~TeV.  
The results of the fits are reported in Table~\ref{tab:fitDBpt}, 
together with the average $\pt$ of D and B mesons in the different 
conditions. The average $\pt$ does not depend strongly on the 
colliding system and on the energy in the centre of mass. On the other 
hand, we remark that $\av{\pt}$ is larger by $\approx 10\%$ 
at mid-rapidity than 
in the forward region ($2.5<y<4$). These two regions correspond 
to the acceptance of the ALICE detector: barrel, $|\eta|<0.9$, 
and forward muon arm, $2.4<\eta<4$.
 
\begin{table}[!t]
  \caption{Parameters derived from the fit of the $\pt$ distributions of 
           D and B mesons to the expression~(\ref{eq:fitMesons}) and average 
           value of $\pt$ for these particles.}
  \label{tab:fitDBpt}
  \begin{center}
  \begin{tabular}{|c|cc|c|c|c|}
\hline
  Particle & System & $\sqrtsNN$ [TeV] & $\pt^0~[\gev/c]$ & $n$ & $\av{\pt} [\gev/c]$\\
\hline
\hline
                    & pp     & 14  & 2.04 & 2.65 & 1.85 \\
                  D & p--Pb  & 8.8 & 2.09 & 2.72 & 1.83 \\
($|y_{\rm lab}|<1$) & Pb--Pb & 5.5 & 2.12 & 2.78 & 1.81 \\
\hline
                      & pp     & 14  & 2.18 & 3.04 & 1.67 \\
                    D & p--Pb  & 8.8 & 2.22 & 3.11 & 1.66 \\
($2.5<y_{\rm lab}<4$) & Pb--Pb & 5.5 & 2.25 & 3.17 & 1.64 \\
\hline
\hline
                    & pp     & 14  & 6.04 & 2.88 & 4.90 \\
                  B & p--Pb  & 8.8 & 6.08 & 2.90 & 4.89 \\
($|y_{\rm lab}|<1$) & Pb--Pb & 5.5 & 6.14 & 2.93 & 4.89 \\
\hline
                      & pp     & 14  & 6.45 & 3.54 & 4.24 \\
                    B & p--Pb  & 8.8 & 6.49 & 3.56 & 4.24 \\
($2.5<y_{\rm lab}<4$) & Pb--Pb & 5.5 & 6.53 & 3.59 & 4.24 \\
\hline
  \end{tabular}
  \end{center}
\end{table}

\section*{Acknowledgments}

The authors would like to acknowledge F.~Antinori, P.~Crochet, M.~Mangano, 
A.~Morsch, G.~Paic, E.~Quercigh and R.~Vogt for many stimulating and fruitful 
discussions.

\section*{Appendix:\\ PYTHIA parameters used for heavy quark generation 
at LHC energies}

In Table~\ref{tab:pythiaparams} we report the complete list of parameters 
used in the PYTHIA event generator~\cite{pythia} in order to reproduce the 
inclusive $\pt$ distribution for the heavy quarks given by the HVQMNR program 
based on NLO calculations by M.~Mangano, P.~Nason and 
G.~Ridolfi~\cite{MNRcode}. 
A detailed description of the parameters
can be found in Ref.~\cite{pythia}.

As specified in Section~\ref{CHAP6_5:generators}, the main parameter 
we tuned is the lower $\pt^{\rm hard}$ limit: the optimal value was 
found to be $2.1~\gev/c$ for charm production and $2.75~\gev/c$ for 
beauty production, both for \mbox{Pb--Pb} collisions at 
$\sqrtsNN=5.5~\tev$ and for pp collisions at $\sqrt{s}=14~\tev$.
Therefore, one can reasonably assume that the same values can be used 
also for \mbox{p--Pb} collisions at $\sqrtsNN=8.8~\tev$.    

The different values for the partonic intrinsic transverse momentum 
$k_{\rm t}$ in pp, \mbox{p--Pb} 
and \mbox{Pb--Pb} collisions were taken from Ref.~\cite{vogtnew}.

\begin{table}[!h]
\caption{PYTHIA parameters used for the generation of charm and
  beauty quarks in pp collisions at 14~TeV, \mbox{p--Pb} collisions at
  8.8~TeV and \mbox{Pb--Pb} collisions at 5.5~TeV. All non-specified parameters
  are left to PYTHIA 6.150 defaults.}
\label{tab:pythiaparams}
\begin{center}
\begin{tabular}{|c|c|c|c|}
\hline
Description & Parameter & Charm & Beauty \\
\hline
\hline
Process types & MSEL & 1 & 1 \\
\hline
Quark mass $[\gev]$ & PMAS(4/5,1) & 1.2 & 4.75 \\
\hline
Minimum $\pt^{\rm hard}~[\gev/c]$ & CKIN(3) & 2.1 & 2.75 \\
\hline
CTEQ 4L     & MSTP(51) & 4032 & 4032 \\
Proton PDF & MSTP(52) & 2 & 2 \\
\hline
Switch off & MSTP(81) & 0 & 0 \\
multiple & PARP(81) & 0 & 0 \\
interactions & PARP(82) & 0 & 0 \\
\hline
Initial- and final-state & MSTP(61) & 1 & 1 \\
parton shower on & MSTP(71) & 1 & 1 \\
\hline
2$^{\mathrm{nd}}$ order $\alpha_s$ & MSTP(2) & 2 & 2 \\
\hline
QCD scales & MSTP(32) & 2 & 2 \\
for hard scattering & PARP(34) & 1 & 1 \\
and parton shower & PARP(67) & 1 & 1 \\
& PARP(71) & 4 & 1 \\
\hline
Intrinsic $k_{\rm t}$ &  &  &  \\
from gaussian distr. with mean 0 & MSTP(91)      &  1    & 1 \\
   $\sigma$ $[\gev/c]$    & PARP(91) & 1.00 (pp) & 1.00 (pp) \\
                &          & 1.16 (p--Pb)   &  1.60 (p--Pb)   \\
                &          & 1.30 (Pb--Pb)   &  2.04 (Pb--Pb)   \\
upper cut-off (at 5 $\sigma$) $[\gev/c]$ & PARP(93) & 5.00 (pp)  & 5.00 (pp) \\
                &          & 5.81 (p--Pb)    & 8.02 (p--Pb)   \\
                &          & 6.52 (Pb--Pb)   &  10.17 (Pb--Pb)   \\
\hline
\end{tabular}
\end{center}
\end{table}

\clearpage
%%%
%%% Bibliography
%%%

\end{document}